\documentclass[%
 reprint,
 amsmath,amssymb,
 aps,
]{revtex4-2}

\usepackage{amsmath,amssymb}
\usepackage{graphicx}
\usepackage{dcolumn}
\usepackage{braket}
\usepackage{bm}
\usepackage{ytableau}
\usepackage[most]{tcolorbox}
\usepackage{algpseudocode}
\usepackage[caption=false]{subfig}

\usepackage[breaklinks=true,colorlinks,citecolor=blue,linkcolor=blue,urlcolor=blue]{hyperref}

\usepackage[capitalise]{cleveref}

\crefname{section}{Sec.}{Secs.}
\Crefname{section}{Section}{Sections}

\newcommand{\Ind}[1]{\mathbf{1}\!\left\{#1\right\}}
\newcommand{\tr}{\mathrm{tr}}

\begin{document}

\addtocontents{toc}{\protect\setcounter{tocdepth}{-10}}

\title{Random Projections for Multi-Copy Quantum Algorithms}

\author{Xiaoyu Liu$^{1,2,*}$}%
\author{Jordi Tura$^{1,2}$}%
\author{Johannes Knörzer$^{3,4,5,\dagger}$}%
\affiliation{$^1$$\langle aQa ^L\rangle $ Applied Quantum Algorithms, Universiteit Leiden}%
\affiliation{$^2$Instituut-Lorentz, Universiteit Leiden, P.O. Box 9506, 2300 RA Leiden, The Netherlands}%
\affiliation{$^3$Department of Physics, ETH Zurich, CH-8093 Zurich, Switzerland}
\affiliation{$^4$Quantum Center, ETH Zurich, CH-8093 Zurich, Switzerland}
\affiliation{$^5$ETH Zurich - PSI Quantum Computing Hub, Paul Scherrer Institute, CH-5232 Villigen, Switzerland}

\date{\today}

\begin{abstract}
Estimating nonlinear properties of quantum states is a central task in quantum information science.
Multivariate traces, $\tr(\rho_1 \cdots \rho_K)$, and nonlinear observables such as $\tr(\rho^K)$, for integer $K$, can be accessed through collective measurements on multiple state copies, but standard protocols based on swap tests require coherent operations on the full Hilbert space and become experimentally unfeasible for large systems.
In this work, we introduce a framework for multi-copy measurements based on random projections onto lower-dimensional subspaces prior to the collective measurement, which is then performed only on the reduced Hilbert space.
This procedure yields a tunable tradeoff between coherent quantum resources and statistical sampling overhead, allowing the amount of coherent processing to be matched to the capabilities of the underlying hardware.
We derive explicit formulas relating the Haar-averaged projected moments to multivariate traces of the original states and analyze the sampling overhead induced by the projection procedure.
Specifically, after compressing an $n$-qubit state to a reduced $q$-qubit subspace, estimating $\tr(\rho^K)$ requires approximately $O(2^{(n-q)(K-1)})$ copies of $\rho$,
with each qubit projected out increasing the sampling cost by a factor of $2^{K-1}$.
Our results establish how coherent multi-copy operations can be traded for additional state copies, enabling multi-copy quantum protocols to be optimized for the available hardware resources.
\end{abstract}
                            
\maketitle

Estimating trace polynomials of quantum states is a fundamental task in quantum information science.
This includes multivariate traces \(\tr(\rho_1 \cdots \rho_K)\)~\cite{quek2024multivariate,fernandes2024unitaryinvariant}, as well as nonlinear state moments $\tr(\rho^K)$, which determine Rényi entropies~\cite{renyi1961measures,muller-lennert2013quantum} and other entanglement measures~\cite{hill1997entanglement, yang2021parametrized, coffman2000distributed, wong2001potential, meyer2002global, beckey2021computable, liu2025generalized, foulds2021controlled, liu2025measuring}.
They appear naturally in a broad class of quantum algorithms~\cite{buhrman2001quantum,ekert2002direct,subramanian2021quantum,wang2023quantum,shin2025resourceefficient}, and play an important role in quantum many-body physics and quantum simulation~\cite{abanin2012measuring,daley2012measuring,linke2018measuring,brydges2019probing,elben2018renyi,vermersch2018unitary,elben2020mixedstate,huang2020predicting, elben2023randomized}.

For integer \(K\), these observables can be accessed with multi-copy measurements acting jointly on \(K\) quantum states.
The backbone of this approach is the swap trick, which states that for two density matrices $\rho_1$ and $\rho_2$, their inner product fulfills \(\tr(\rho_1 \rho_2) = \tr(\mathbb{S}\,(\rho_1\otimes\rho_2)),\) where \(\mathbb{S}\) denotes the swap operator exchanging the two quantum states.
More generally, the identity \( \tr(\rho_1 \cdots \rho_K) = \tr(V_K\,(\rho_1\otimes\cdots\otimes\rho_K))\) holds,
with \(V_K\) the cyclic permutation operator acting on the states $\rho_1, \cdots, \rho_K$.
These well-known facts underlie a broad family of multi-copy quantum protocols realizing generalized swap~\cite{brun2004measuring, cotler2019quantum, huggins2021virtual, koczor2021exponential, johri2017entanglement} and permutation tests~\cite{barenco1997stabilization, buhrman2026permutation, kada2008efficiency, laborde2024quantum}.
In many settings, coherent access to multiple copies yields exponentially improved sample complexity compared to protocols restricted to single-copy measurements~\cite{haah2017sampleoptimal, acharya2020estimating, wang2023quantum}.
Conversely, estimating trace polynomials from only local or single-copy measurements can become exponentially costly in the system size~\cite{ye2025exponential, anshu2022distributed}.

At the same time, realizing coherent multi-copy interference is challenging:
for an underlying Hilbert space of dimension $d$, these operations act on a space of dimension $d^K$ and generally require large-scale entangling operations or coherent quantum communication between different subsystems or devices~\cite{knorzer2026distributed}.
As the system size and number of copies increase, these requirements rapidly become prohibitive in implementations.

This naturally raises the question whether there exists an intermediate regime between fully coherent multi-copy protocols, which is experimentally demanding, and completely local measurement schemes, whose sample complexity can scale exponentially with the system size.
Recent work on distributed inner product estimation has provided one possible answer in the two-copy setting:
estimating inner products $\mathrm{tr}(\rho_1 \rho_2)$ up to additive error $\varepsilon$ using only local operations and classical communication requires \(\mathcal{O}(\max\{1/\varepsilon^2,\sqrt{d}/\varepsilon\})\) copies in dimension \(d\)~\cite{anshu2022distributed},
whereas allowing a limited amount of coherent quantum processing leads to tunable tradeoffs between sample complexity and coherent control requirements~\cite{arunachalam2024distributed}.
The latter approach combines randomization with swap tests on lower-dimensional projected states, drawing inspiration from low-distortion embeddings in high-dimensional geometry.
However, it is restricted to two-copy observables and the estimation of pure-state overlaps \(|\braket{\psi|\phi}|^2\).

\begin{figure*}
  \includegraphics[width=\linewidth]{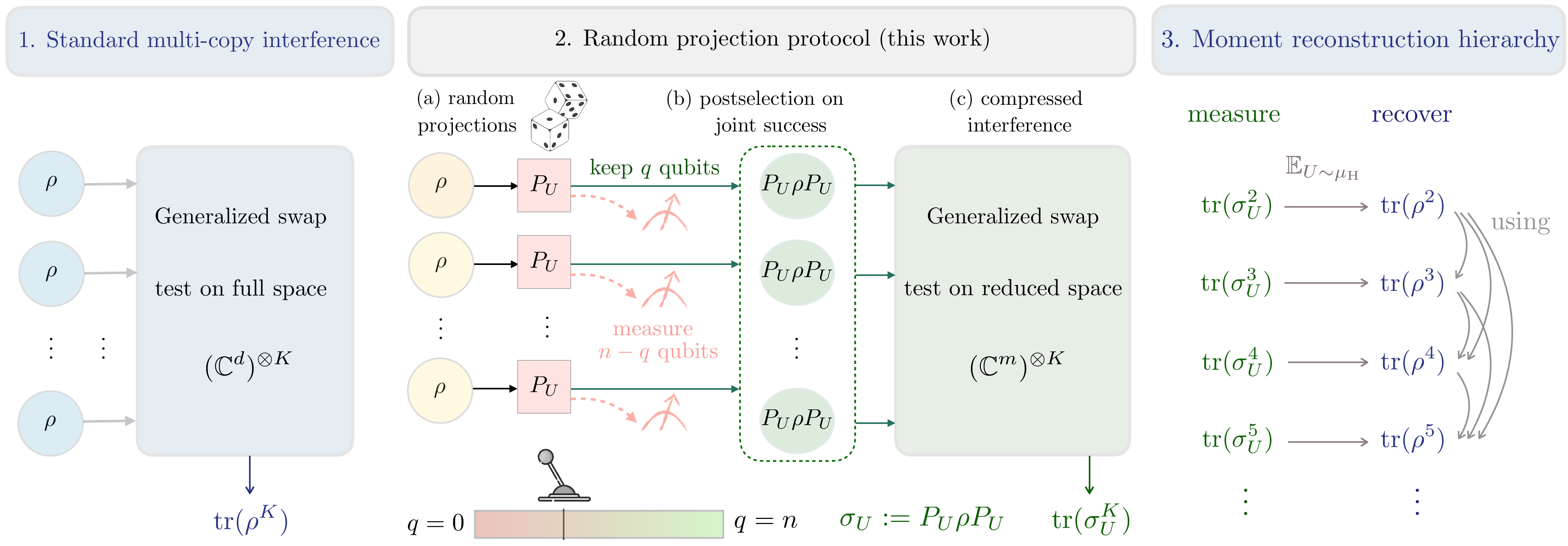}
  \caption{
    Comparison between standard multi-copy protocols and the random projection framework introduced in this work.
    \emph{Left:} Conventional estimation of nonlinear observables such as $\tr(\rho^K)$ using generalized swap test acting on the full Hilbert space $(\mathbb C^d)^{\otimes K}$.
    \emph{Center:} Random projection protocol.
    (a) Each copy of the quantum state $\rho$ is projected onto the same Haar-random $m$-dimensional subspace ($m< d$) using the projector $P_U = U P U^\dagger$, where $m=2^q$ for $q$ qubits.
    (b) Runs are postselected on joint projection success, yielding projected states $P_U \rho P_U$.
    (c) The generalized swap test is subsequently performed only on the reduced Hilbert space $(\mathbb C^m)^{\otimes K}$.
    \emph{Right:} Schematic hierarchy of moment reconstruction relations.
    The projected moments $\mathrm{tr}(\sigma_U^K)$ determine the target moments $\mathrm{tr}(\rho^K)$ recursively through coupled lower-order terms.
    }
  \label{fig:schematic-figure}
\end{figure*}

More broadly, randomization methods have become an increasingly powerful tool throughout quantum information science.
Randomized measurements and classical-shadow protocols have enabled efficient estimation of many properties of quantum systems using random local unitaries and statistical post-processing~\cite{elben2020mixedstate, elben2023randomized, huang2020predicting,elben2018renyi,vermersch2018unitary,brydges2019probing,li2026quantum}.
These developments have shown that randomness can substantially reduce resource requirements in a variety of information processing and learning tasks.

In this work, we introduce a protocol for estimating multivariate traces of quantum states, including state moments as a special case, based on random projections onto lower-dimensional subspaces.
Instead of performing generalized swap tests directly on the original Hilbert space ($d=2^n$), we first project each state copy onto a lower-dimensional subspace ($m=2^q$) and subsequently perform the collective measurement only on the reduced space.
We show that the required number of state copies scales approximately as $O(2^{(n-q)(K-1)})$.
The resulting framework yields a tunable tradeoff between coherent quantum resources and classical post-processing.

\emph{Random projection.}\textemdash
The multivariate trace \(\tr(\rho_1 \cdots \rho_K)\) can be estimated with a generalized swap test as briefly reviewed in \cref{sec:preliminaries}.
In our protocol, instead of performing the generalized swap test directly on $(\mathbb C^d)^{\otimes K}$, each of the state copies $\rho_1,\cdots,\rho_K$ is first projected onto a random rank-$m$ subspace ($m< d$).
A central quantity of the protocol is the Haar-averaged
projected observable
\begin{equation}\label{eq:definition-sigma}
\bar \sigma^{(d,m)}_K(\rho_1,\cdots,\rho_K) := \mathbb E_{U\sim\mu_{\rm H}}
\!\left[
\tr(\sigma_{1,U} \cdots \sigma_{K,U})
\right],
\end{equation}
with \( \sigma_{i,U} = P_U \rho_i P_U \ (i=1,\cdots,K)\) and \( P_U = UPU^\dagger \).
\( P \) is a fixed rank-$m$ projector and \( U\sim\mu_{\rm H} \) is drawn from the Haar measure on the unitary group \( \mathrm U(d) \).\\

\textbf{Theorem 1.}
\textit{Let $\rho_1, \cdots, \rho_K$ be density matrices on $\mathbb C^d$, and let $S_K$ denote the symmetric group on $K$ elements, with $\pi=(1\,2\,\cdots\,K)\in S_K$ the cyclic permutation.
Then
\begin{equation}
\bar \sigma^{(d,m)}_K(\rho_1, \cdots, \rho_K)
=
\sum_{\tau\in S_K}
\gamma_\tau(d,m)
\prod_{j=1}^{\mathrm{c}(\tau\pi)}
\tr\!\left ( \prod_{\ell\in C_j(\tau\pi)} \rho_\ell \right ),
\label{eq:general_projected_multivariate_main}
\end{equation}
where
\begin{equation}
\gamma_\tau(d,m)
=
\sum_{\alpha\in S_K}
m^{\mathrm{c}(\alpha)}
\,\mathrm{Wg}_d(\alpha^{-1}\tau),
\label{eq:c_lambda_formula}
\end{equation}
with $\mathrm{c}(\alpha)$ the number of cycles in $\alpha\in S_K$, $C_j(\alpha)$ its $j$th cycle, whose elements are multiplied in cyclic order, and $\mathrm{Wg}_d$ the unitary Weingarten function.}

A proof is given in~\cref{sec:supp-proof-theorem1}.
Theorem~1 expresses the Haar-averaged projected observables as universal combinations of multivariate trace invariants of the original states.
In general, different cyclic orderings are coupled through \cref{eq:general_projected_multivariate_main}.
For example, for \(K=3\), the quantities \( \tr(\rho_1\rho_2\rho_3)\) and \( \tr(\rho_1\rho_3\rho_2)\) appear together and can be separated by combining projected observables corresponding to different cyclic orderings.
Since these quantities are generally complex, both their real and imaginary parts must be estimated.
An explicit example for $K=3$ is discussed in the End Matter.
In the following, we focus on the case $\rho_1 = \cdots = \rho_K$ for its simplicity and importance.\\

\textbf{Corollary 1.}
\textit{Let \(\rho\) be a density matrix on \(\mathbb{C}^d\).
Then
\begin{equation}
\bar\sigma^{(d,m)}_K(\rho) = 
\sum_{\tau\in S_K}
\gamma_\tau(d,m)
\prod_{j=1}^{\mathrm{c}(\tau\pi)}
\tr\!\left ( \rho^{\nu_j} \right ),
\label{eq:general_projected_moments_main}
\end{equation}
where \( \bar \sigma^{(d,m)}_K(\rho) \equiv \bar \sigma^{(d,m)}_K(\rho, \cdots, \rho) \), \(\gamma_\tau\) as in \cref{eq:c_lambda_formula}, and \(\nu_j\) denotes the cycle length of the $j$th cycle.
}\\

Corollary~1 shows that the Haar-averaged projected moments are universal combinations of the moments $\tr(\rho^2),\cdots,\tr(\rho^K)$, with coefficients determined solely by the projection rank $m$, the
Hilbert-space dimension $d$, and the permutation structure of $S_K$.
Fig.~\ref{fig:schematic-figure} illustrates the random projection protocol for this case.
Note that Theorem~1 and Corollary~1 concern the expectation value in \cref{eq:definition-sigma}, while the variance formula of projected moments is provided in \cref{sec:variance-moments}.

\emph{Expansion coefficients}.\textemdash
The coefficients \(\gamma_\tau(d,m)\) admit a particularly convenient
representation in terms of irreducible characters of the symmetric
group \(S_K\).
Using Weingarten calculus~\cite{kostenberger2021weingarten}, which comprises a collection of techniques for evaluating averages of polynomials of matrix elements of Haar-random unitary matrices, the expansion coefficients can be expressed as
\begin{equation}\label{eq:coefficients-weingarten}
\gamma_\tau(d,m)
=
\frac1{K!}
\sum_{\substack{\lambda\vdash K,\\ \ell(\lambda)\leqslant d}}
f^\lambda
\frac{\alpha_\lambda(m)}{\alpha_\lambda(d)}
\chi_\lambda(\tau),
\end{equation}
where the sum runs over integer partitions \(\lambda\vdash K\) with at most \(d\) rows, denoted \(\ell(\lambda)\leqslant d\), \(f^\lambda = \dim(\lambda)\) is the dimension of the corresponding irrep, and \(\chi_\lambda(\tau)\) its irreducible character evaluated for the permutation $\tau \in S_K$.
Lastly, \(\alpha_\lambda\)
denotes the content polynomial of $\lambda$~\cite{macdonald1995symmetric}.
Details on the derivation of Eq.~\eqref{eq:coefficients-weingarten} are given in~\cref{sec:supp-proof-theorem1} of the Supplemental Material, and examples for $K\in\{2,3,4\}$ are provided in~\ref{app:low-order-examples}.

\emph{Moment reconstruction hierarchy.}\textemdash
Eqs.~\eqref{eq:general_projected_moments_main} and
\eqref{eq:coefficients-weingarten}
define a hierarchy of projected moment relations connecting the Haar-averaged observables
$\bar\sigma_K$ to the spectral moments
$p_K=\tr(\rho^K)$.
The moment $p_K$ arises exclusively from permutations whose cycle structure consists of a single $K$-cycle, while all other cycle structures generate monomials involving only moments $p_j$ with $j<K$.
Collecting all contributions proportional to $p_K$ into a coefficient $\gamma^{(K)}(d,m)$ therefore yields
\begin{equation}
    \bar \sigma^{(d,m)}_K
    =
    \gamma^{(K)}(d,m)\, p_K
    +
    F_K\!\left(p_1,\cdots,p_{K-1}\right),
\end{equation}
where $F_K$ is a polynomial in lower-order moments.
Solving for the highest-order moment gives
\begin{equation}
    p_K
    =
    \frac{
    \bar \sigma^{(d,m)}_K
    -
    F_K\!\left(p_1,\cdots,p_{K-1}\right)
    }{
    \gamma^{(K)}(d,m)
    }.
\end{equation}
This recursive relation defines the moment reconstruction hierarchy depicted in Fig.~\ref{fig:schematic-figure}.

The simplest nontrivial case is $K=2$, corresponding to the purity $\tr(\rho^2)$.
A direct calculation gives
\begin{equation}
\bar \sigma^{(d,m)}_2
=
\frac{dm^2-m}{d(d^2-1)} \ p_2+\frac{dm-m^2}{d(d^2-1)},
\label{eq:purity_formula}
\end{equation}
and as such, the purity is recovered directly from the Haar-averaged projected moment.
For $K=3$, the projected third moment takes the form
\begin{equation}
\bar \sigma^{(d,m)}_3
=
\gamma^{(3)}(d,m)~p_3
+
\alpha(d,m)~p_2
+
\beta(d,m),
\label{eq:third_moment_structure}
\end{equation}
where the explicit results for $\alpha(d,m)$, $\beta(d,m)$ and $\gamma^{(3)}(d,m)$ are provided in~\cref{app:low-order-examples}.
This illustrates that recovering
\(p_K\) with \(K\geqslant 3\) requires knowledge of the lower-order moments
\(p_2, \cdots, p_{K-1}\) in addition to $\bar \sigma_K$.

\emph{Subsystem implementation.}\textemdash
For $d=2^n$ and $m=2^q$, the random projection protocol admits a natural operational realization in terms of subsystem measurements.
After applying a random unitary $U$ to each copy, $n-q$ qubits are measured in the computational basis.
One may postselect (i) on all measured qubits being found in the state $|0\rangle$, or, more generally, (ii) accept any outcome provided all $K$ copies yield the same measurement result.
In either case, the remaining $q$ qubits are restricted to a common $m$-dimensional subspace on which the generalized swap test is subsequently performed.

The latter strategy (ii) effectively sums over all
$L=d/m$ equivalent projection sectors.
Since the corresponding Haar-twirled observable is larger by a factor of $L$, it yields the same moment reconstruction formulas after multiplication by the known normalization factor $L^{-1}$, while increasing the postselection success probability by a factor of approximately $L$.
Throughout this work, we employ this strategy in our numerical studies, and the corresponding explicit algorithm is given in~\cref{sec:sm-sample-complexity}.
Note that more sample-efficient postselection strategies are possible with quantum memories, which may avoid some of the overhead associated with probabilistic projections~\cite{arunachalam2024distributed}.
Yet we restrict ourselves to the memory-free setting, motivated by the prospect of near-term implementations.

\begin{figure}[b]
  \includegraphics[width=\linewidth]{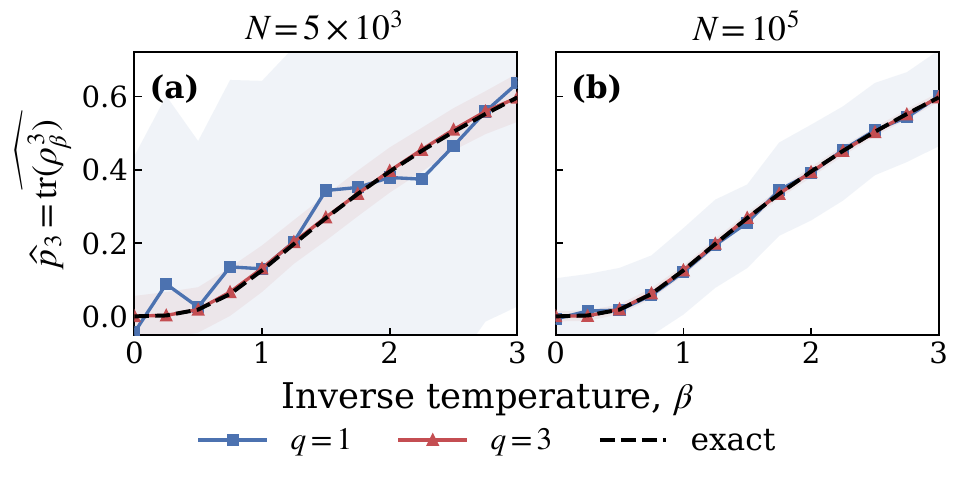}
  \caption{
    Estimated third moment $\hat p_3$ as a function of inverse temperature $\beta$ for thermal states of the five-qubit transverse-field Ising model from \cref{eq:ising-model} with $J=h=1$.
    Results are shown for two total numbers of copies consumed by the protocol: (a) $N=5\times 10^3$ and (b) $N=10^5$. The dashed black curve shows the exact value $p_3=\mathrm{tr}(\rho_\beta^3)$.
    Colored markers denote reconstructed estimates obtained using random projections that retain $q=1$ (blue squares) and $q=3$ (red triangles) qubits prior to collective measurement.
    Shaded regions indicate interquartile ranges obtained from $500$ independent realizations of the complete estimation protocol.
    }
  \label{fig:thermal-state-rho3}
\end{figure}

\emph{Thermal state example}.\textemdash
As an example, we illustrate the reconstructed third moment,
$p_3=\tr(\rho_\beta^3)$, for thermal states
\(\rho_\beta=e^{-\beta H}/\mathbf{Z}\) of the transverse-field Ising model,
\begin{equation}\label{eq:ising-model}
H=-J\sum_{i=1}^{n-1} Z_iZ_{i+1}-h\sum_{i=1}^n X_i,
\end{equation}
as a function of inverse temperature $\beta$, with partition function
\(\mathbf{Z}=\tr(e^{-\beta H})\).
Fig.~\ref{fig:thermal-state-rho3}
shows the estimated third moment $\hat p_3$ as a function of inverse temperature $\beta$ for different projection ranks and numbers of consumed state copies $N$, for a system size \(n=5\).
These density-matrix simulations are carried out with random projections sampled from the Haar measure.
For each protocol execution, a random projected moment is generated and the corresponding postselection and generalized swap test outcomes are sampled according to their exact probabilities.
The reconstruction is performed recursively from finite-sample estimates of both $\hat p_2$ and $\hat p_3$, i.e., no prior knowledge of the purity of $\rho_\beta$ is assumed.

We use the same number of protocol executions to estimate each $\hat{p}_K$.
For example, in Fig.~\ref{fig:thermal-state-rho3}(a), we use $10^3$ executions of the 2-copy protocol to estimate $\hat{p}_2$ and $10^3$ executions of the 3-copy protocol to estimate $\hat{p}_3$, consuming a total of $5\times 10^3$ copies of $\rho_{\beta}$.
As expected, increasing the number of used state copies $N$ systematically reduces statistical fluctuations and improves agreement with the exact result.
For a fixed total number of state copies, increasing the projection rank \(m\), or equivalently the number of coherently processed qubits \(q\), leads to progressively better agreement with the exact result, which illustrates the tradeoff between coherent quantum resources and statistical sampling overhead.

\begin{figure}
  \includegraphics[width=0.99\linewidth]{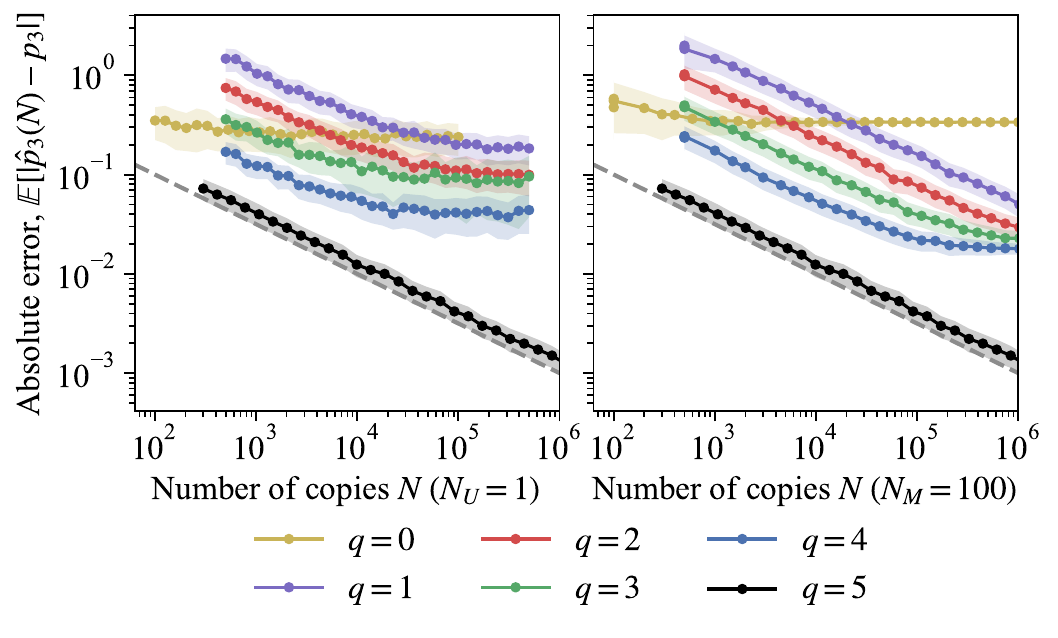}
  \caption{
    Mean absolute estimation error $\mathbb{E}[|\hat p_K(N)-p_K|]$ for the reconstructed third moment $p_3=\mathrm{tr}(\rho^3)$ as a function of the number of consumed state copies $N$, for different numbers $q$ of kept qubits.
    The left panel fixes the number of randomized unitaries to $N_U=1$ and varies the number of measurement shots $N_M$, while the right panel fixes $N_M=100$ and varies $N_U$.
    The target state is a noisy 5-qubit GHZ state with \(30\%\) depolarizing noise.
    Solid curves show averages over 500 independent realizations, while shaded regions indicate $\pm 1/3$ standard deviation.
    The dashed guide indicates the expected Monte Carlo scaling $\propto N^{-1/2}$.
    The fully coherent case, $q=5$, corresponds to the generalized swap test, while the fully local case, $q=0$, follows the protocol in~\cite{li2026quantum}.
    }
  \label{fig:mean-error-K3}
\end{figure}

\emph{Approximate random projections.}\textemdash
Implementing exact Haar-random unitaries on large systems is generally infeasible.
In practice, random projections may instead be generated using shallow random circuits that approximate low-order unitary designs~\cite{brandao2016local, haferkamp2022random, schuster2025random, cui2025unitary}.
In particular, the projected $K$th moments are determined by Haar averages of polynomials that are degree $K$ in both the matrix elements of $U$ and those of $U^\dagger$.
A unitary $K$-design is defined precisely by its ability to reproduce such moments of the Haar measure.
Consequently, approximate unitary $K$-designs provide a natural route toward realizing the random projections required by the protocol.
To investigate this setting, we replace Haar-random unitaries by depth-5 brickwork circuits composed of alternating layers of nearest-neighbor two-qubit gates~\cite{brandao2016local, haferkamp2022random}.

\emph{Sample complexity.}\textemdash 
We now discuss how the reconstruction error depends on the two sampling parameters of the protocol.
We denote by \(N_U\) the number of sampled random unitaries and by \(N_M\) the number of measurements performed per unitary. 
In the hierarchical setting, estimating $p_K$ requires estimating the lower-order moments $p_2,\cdots,p_{K-1}$.
Assuming the same values of $N_U$ and $N_M$ are used for each order, the total number of copies required up to order $K$ is $N=\frac{(K+2)(K-1)}{2}N_UN_M$.
For example, in Fig.~\ref{fig:thermal-state-rho3}, we vary \(N_U\) while fixing \(N_M=1\), so that the total number of copies used to estimate \(p_3\) is \(N=5N_U\).

The sample complexity is summarized by the following observation.

\textbf{Observation 1.}
\textit{
Let $\epsilon_p = |\hat{p}_K - p_K|$ and denote the brickwork approximation error by $\epsilon_{\mathrm{bw}}=|\mathbb{E}_{\mathrm{brickwork}}[\hat{p}_K]-p_K|$ where $\mathbb{E}_{\mathrm{brickwork}}[\hat{p}_K]$ is the infinite sampling limit of the full hierarchical estimator under the brickwork ensemble. 
Then:
\begin{equation}
\begin{split}
    \epsilon_p \sim O\left( \mathrm{poly}(K)\sqrt{\frac{1}{N_U}+\frac{2^{(n-q)(K-1)}}{N_U N_M}}  \right) + \epsilon_{\mathrm{bw}}.   
\end{split}
\label{eq:complexity}
\end{equation}
}\\
This observation applies to the hierarchical setting, where the lower-order moments are themselves estimated rather than assumed known.
The factor $\mathrm{poly}(K)$ accounts for error propagation from the recursively estimated lower-order moments when the same $N_U$ and $N_M$ are used for all orders.
For small values of $K$, the corresponding accumulated amplification factor remains moderate.
The bound consists of two contributions, the statistical error and $\epsilon_{\mathrm{bw}}$.
The statistical error itself contains two sources:
the first term in the square root arises from sampling a finite number of random unitaries, while the second comes from the finite number of measurements per unitary.
Increasing $N_U$ reduces both statistical contributions, whereas increasing $N_M$ only reduces the second one.
This explains why increasing $N_U$ is sufficient to reduce $\epsilon_p$ even when $N_M=1$, as shown in Fig.~\ref{fig:thermal-state-rho3}.

Fig.~\ref{fig:mean-error-K3} further illustrates the effects of choosing different values of $N_U$ and $N_M$.
As an example, we consider the estimation of the third moment of a noisy GHZ state $\rho_\mathrm{GHZ}(n,w)=(1-w)\ket{\mathrm{GHZ}_n}\bra{\mathrm{GHZ}_n}+2^{-n}w\mathbb{I}$ with $\ket{\mathrm{GHZ}_n} = 2^{-1/2}(\ket{0}^{\otimes n} + \ket{1}^{\otimes n})$.
We set $n=5$ and $w=0.3$.
For all compression levels, the estimation error initially decreases approximately as $O(N^{-1/2})$.
When $N_U$ is limited, the error may plateau because of the finite-unitary sampling contribution.
Even when both $N_U$ and $N_M$ are sufficiently large, the error can still saturate because of the brickwork approximation error $\epsilon_{\mathrm{bw}}$.
The two limiting cases of \(q\) recover two known protocols:
when \(q=5\), our protocol reduces to the generalized swap test, whereas when \(q=0\), it recovers the very recent fully local randomized protocol in~\cite{li2026quantum}. 
The latter performs better in the small-copy regime than some of the intermediate \(q\), but its error saturates earlier and at a larger value.
Further comparisons between our protocol and the fully local protocol are provided in~\cref{sec:q0-comparison}.

\begin{figure}
  \includegraphics[width=0.99\linewidth]
  {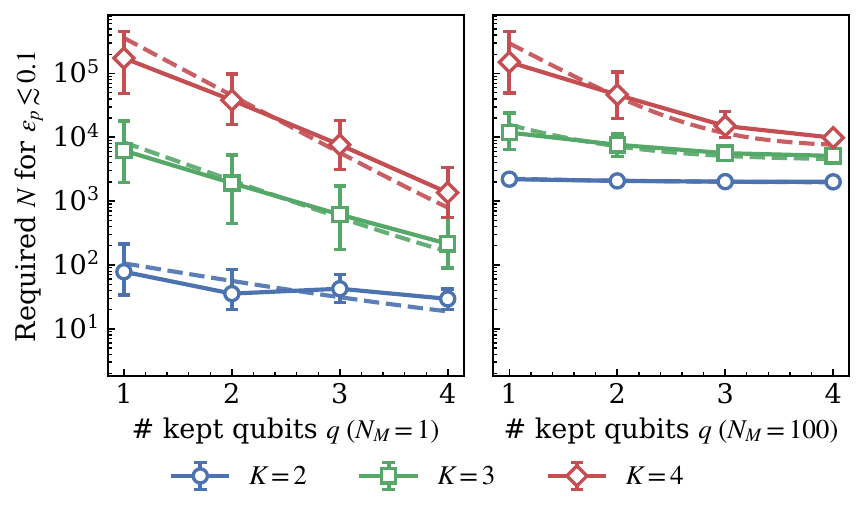}
  \caption{
  Average required number of copies $N$ to obtain $\epsilon_p\lesssim0.1$ for 5-qubit GHZ state with $30\%$ depolarizing noise over 300 independent trials. 
  The error bars indicate the middle 68\% range of the results.
  The brickwork circuit depth is fixed to be 5.
  We fix $N_M=1$ (left panel) and $N_M=100$ (right panel) respectively.
  Higher-order moments are estimated hierarchically by first estimating the lower-order moments, 
  and the required number of copies is counted in total.
  The results are fitted to $O\left(\mathrm{poly}(K)\frac{N_M+2^{(n-q)(K-1)}}{(\epsilon_p - \epsilon_{\mathrm{bw}})^{2}}\right)$.
  The required $N$ decreases as $q$ increases.
  Also, increasing $N_U$ plays a more important role in reducing $\epsilon_p$ than increasing $N_M$, consistent with Observation 1.
  }
  \label{fig:sampling-complexity}
\end{figure}

From Observation 1, the total number of copies then scales as $N \sim O\left(\mathrm{poly}(K)\frac{N_M+2^{(n-q)(K-1)}}{(\epsilon_p - \epsilon_{\mathrm{bw}})^{2}}\right)$, which is dominated by $\sim2^{(n-q)(K-1)}$ when $N_M$ is not too large.
Thus, the required number of copies decreases as $q$ increases.
Fig.~\ref{fig:sampling-complexity} shows the average required $N$ to achieve $\epsilon_p\lesssim0.1$ for $N_M=1$ and $N_M=100$, respectively, together with the corresponding theoretical complexity fits.
A detailed derivation of this scaling is given in~\cref{sec:sm-sample-complexity}.

\emph{Partial transpose}.\textemdash
Our framework naturally extends to moments of partially transposed density matrices.
For a bipartite state \(\rho_{AB}\), let \( \rho_{AB}^{T_B}\) denote the partial transpose with respect to subsystem \(B\).
The moments \(\tr((\rho_{AB}^{T_B})^K)\) admit the representation
\begin{equation}
    \tr\left( (\rho_{AB}^{T_B})^K \right) = \tr\left ( \rho_{AB}^{\otimes K} \left( V_K^{(A)} \otimes [V_K^{(B)}]^{-1}\right) \right),
\end{equation}
where \(V_K^{(A)}\) and \(V_K^{(B)}\) denote cyclic permutations of \(K\) copies on subsystems \(A\) and \(B\), respectively.
Thus, the partial transpose corresponds to reversing the cyclic ordering on one of the two subsystems.
Consequently, moments of the partial transpose can be treated within the same random projection framework by replacing the global permutation operator with the corresponding subsystem permutation operators.

The projected observables are then linear combinations of a family of subsystem-permutation invariants
\begin{equation}
    I_{r,s} = \tr\left( \rho_{AB}^{\otimes K} \left(V_r^{(A)} \otimes V_s^{(B)}\right) \right),\qquad r,s\in S_K,
\end{equation}
where \(V_r^{(A)}\) and  \(V_s^{(B)}\) denote permutation operators associated with \(r,s \in S_K\) acting on the \(A\) and \(B\) subsystems, respectively.
Collecting these in a vector \(\mathbf{I} = (I_{r,s})\), and the projected observables in a vector \(\mathbf{M} = (M_{r,s})\), the random projection protocol yields a linear system
\begin{equation}
    \mathbf{M} = C \mathbf{I},
\end{equation}
where the matrix \(C\) is determined entirely by the subsystem dimensions and projection ranks.
Reconstruction of \(\tr((\rho_{AB}^{T_B})^K)\) can be achieved by inverting this linear system and extracting \(I_{\pi,\pi^{-1}}\), where $\pi=(1\,2\,\cdots\,K)$, as shown in the End Matter.
Note that the quantum part of the framework remains unchanged, as one still performs generalized swap tests on \(K\) copies after projection.
The additional complexity enters only through the enlarged classical reconstruction problem.

\emph{Conclusion \& Outlook}.\textemdash
In this work, we have introduced a random projection framework for estimating multivariate traces of quantum states, including nonlinear quantities such as \(\tr(\rho^K)\) and moments of the partial transpose of \(\rho\).
By projecting quantum states onto low-dimensional subspaces prior to a swap test, the protocol allows for an interpolation between fully coherent multi-copy measurements and fully local randomized schemes.
An interesting direction for future work is an extension to a broader class of observables \(\tr(O\rho^K)\) and a detailed analysis of the associated sample complexities.

We derived explicit reconstruction formulas relating averaged projected observables to the target quantity and showed that the resulting estimators satisfy a hierarchy of coupled equations.
Numerical simulations demonstrate how the protocol performs under substantial dimensionality reduction and using shallow random circuits.
The corresponding sample complexity, in terms of the required number of state copies, scales approximately as $O(2^{(n-q)(K-1)})$, showing explicitly how increasing the compressed subsystem size $q$ reduces the statistical overhead.

By replacing collective measurements on the full Hilbert space with generalized swap tests on reduced subspaces, random projections provide a possible route toward realizing multi-copy algorithms on near-term quantum hardware.
Understanding the resulting tradeoffs between experimentally demanding coherent quantum resources, circuit complexity and statistical sampling overhead is a promising direction for future work.
It will also be interesting to tailor these randomized algorithms to hardware-specific constraints and noise models.

\textit{Acknowledgments}.\textemdash
We thank I.~Cirac and A.~Elben for their valuable insights.
J.K.~acknowledges financial support by the Swiss State Secretariat for Education, Research and Innovation under contract number UeM019-11, and by ETH Zurich.
J.T. acknowledges the support received from the European Union's Horizon Europe research and innovation programme through the ERC StG FINE-TEA-SQUAD (Grant No. 101040729). J.T. also acknowledges the support received by the Dutch National Growth Fund (NGF), as part of the Quantum Delta NL programme. This work is part of the "Quantum Inspire the Dutch Quantum Computer in the Cloud" project (with project number [NWA.1292.19.194]) of the NWA research programme "Research on Routes by Consortia (ORC)," which is funded by the Netherlands Organization for Scientific Research (NWO).

The views and opinions expressed here are solely those of the authors and do not necessarily reflect those of the funding institutions. Neither of the funding institutions can be held responsible for them.

\noindent\rule{0.35\columnwidth}{0.4pt}

\smallskip

\noindent
{\footnotesize
\begin{tabular}{@{}ll}
$^*$ & \normalfont\href{mailto:	xiaoyu@lorentz.leidenuniv.nl}{xiaoyu@lorentz.leidenuniv.nl}\\
$^\dagger$ & \normalfont\href{mailto:jknoerzer@ethz.ch}{jknoerzer@ethz.ch}
\end{tabular}
}
\bigskip
\bibliography{random-projection}

@article{abanin2012measuring,
  title = {Measuring {{Entanglement Entropy}} of a {{Generic Many-Body System}} with a {{Quantum Switch}}},
  author = {Abanin, Dmitry A. and Demler, Eugene},
  year = 2012,
  month = jul,
  journal = {Physical Review Letters},
  volume = {109},
  number = {2},
  pages = {020504},
  issn = {0031-9007, 1079-7114},
  doi = {10.1103/PhysRevLett.109.020504},
  urldate = {2026-05-20},
  copyright = {http://link.aps.org/licenses/aps-default-license},
  langid = {english}
}

@article{acharya2020estimating,
  title = {Estimating {{Quantum Entropy}}},
  author = {Acharya, Jayadev and Issa, Ibrahim and Shende, Nirmal V. and Wagner, Aaron B.},
  year = 2020,
  month = aug,
  journal = {IEEE Journal on Selected Areas in Information Theory},
  volume = {1},
  number = {2},
  pages = {454--468},
  issn = {2641-8770},
  doi = {10.1109/JSAIT.2020.3015235},
  urldate = {2026-05-22},
  copyright = {https://ieeexplore.ieee.org/Xplorehelp/downloads/license-information/IEEE.html}
}

@inproceedings{anshu2022distributed,
  title = {Distributed Quantum Inner Product Estimation},
  booktitle = {Proceedings of the 54th {{Annual ACM SIGACT Symposium}} on {{Theory}} of {{Computing}}},
  author = {Anshu, Anurag and Landau, Zeph and Liu, Yunchao},
  year = 2022,
  month = jun,
  eprint = {2111.03273},
  primaryclass = {quant-ph},
  pages = {44--51},
  doi = {10.1145/3519935.3519974},
  urldate = {2026-06-15},
  archiveprefix = {arXiv}
}

@misc{arunachalam2024distributed,
  title = {Distributed Inner Product Estimation with Limited Quantum Communication},
  author = {Arunachalam, Srinivasan and Schatzki, Louis},
  year = 2024,
  month = oct,
  number = {arXiv:2410.12684},
  eprint = {2410.12684},
  primaryclass = {quant-ph},
  publisher = {arXiv},
  doi = {10.48550/arXiv.2410.12684},
  urldate = {2026-06-16},
  archiveprefix = {arXiv}
}

@article{barenco1997stabilization,
  title = {Stabilization of {{Quantum Computations}} by {{Symmetrization}}},
  author = {Barenco, Adriano and Berthiaume, Andr{\'e} and Deutsch, David and Ekert, Artur and Jozsa, Richard and Macchiavello, Chiara},
  year = 1997,
  month = oct,
  journal = {SIAM Journal on Computing},
  volume = {26},
  number = {5},
  pages = {1541--1557},
  issn = {0097-5397},
  doi = {10.1137/S0097539796302452},
  urldate = {2026-05-27}
}

@article{beckey2021computable,
  title = {Computable and {{Operationally Meaningful Multipartite Entanglement Measures}}},
  author = {Beckey, Jacob L. and Gigena, N. and Coles, Patrick J. and Cerezo, M.},
  year = 2021,
  month = sep,
  journal = {Physical Review Letters},
  volume = {127},
  number = {14},
  pages = {140501},
  publisher = {American Physical Society},
  doi = {10.1103/PhysRevLett.127.140501},
  urldate = {2026-05-26}
}

@article{brandao2016local,
  title = {Local {{Random Quantum Circuits}} Are {{Approximate Polynomial-Designs}}},
  author = {Brand{\~a}o, Fernando G. S. L. and Harrow, Aram W. and Horodecki, Micha{\l}},
  year = 2016,
  month = sep,
  journal = {Communications in Mathematical Physics},
  volume = {346},
  number = {2},
  pages = {397--434},
  issn = {1432-0916},
  doi = {10.1007/s00220-016-2706-8},
  urldate = {2026-06-12},
  langid = {english}
}

@article{brun2004measuring,
  title = {Measuring Polynomial Functions of States},
  author = {Brun, T.A.},
  year = 2004,
  month = sep,
  journal = {Quantum Information and Computation},
  volume = {4},
  number = {5},
  pages = {401--408},
  issn = {15337146, 15337146},
  doi = {10.26421/QIC4.5-6},
  urldate = {2026-06-15}
}

@article{brydges2019probing,
  title = {Probing {{R\'enyi}} Entanglement Entropy via Randomized Measurements},
  author = {Brydges, Tiff and Elben, Andreas and Jurcevic, Petar and Vermersch, Beno{\^i}t and Maier, Christine and Lanyon, Ben P. and Zoller, Peter and Blatt, Rainer and Roos, Christian F.},
  year = 2019,
  month = apr,
  journal = {Science},
  volume = {364},
  number = {6437},
  pages = {260--263},
  publisher = {American Association for the Advancement of Science},
  doi = {10.1126/science.aau4963},
  urldate = {2026-06-15}
}

@article{buhrman2001quantum,
  title = {Quantum {{Fingerprinting}}},
  author = {Buhrman, Harry and Cleve, Richard and Watrous, John and {de Wolf}, Ronald},
  year = 2001,
  month = sep,
  journal = {Physical Review Letters},
  volume = {87},
  number = {16},
  pages = {167902},
  publisher = {American Physical Society},
  doi = {10.1103/PhysRevLett.87.167902},
  urldate = {2026-06-15}
}

@misc{buhrman2026permutation,
  title = {Permutation Tests for Quantum State Identity},
  author = {Buhrman, Harry and Grinko, Dmitry and Lunel, Philip Verduyn and Weggemans, Jordi},
  year = 2026,
  month = apr,
  number = {arXiv:2405.09626},
  eprint = {2405.09626},
  primaryclass = {quant-ph},
  publisher = {arXiv},
  doi = {10.48550/arXiv.2405.09626},
  urldate = {2026-05-27},
  archiveprefix = {arXiv}
}

@article{coffman2000distributed,
  title = {Distributed Entanglement},
  author = {Coffman, Valerie and Kundu, Joydip and Wootters, William K.},
  year = 2000,
  month = apr,
  journal = {Physical Review A},
  volume = {61},
  number = {5},
  pages = {052306},
  publisher = {American Physical Society},
  doi = {10.1103/PhysRevA.61.052306},
  urldate = {2026-05-26}
}

@article{cotler2019quantum,
  title = {Quantum {{Virtual Cooling}}},
  author = {Cotler, Jordan and Choi, Soonwon and Lukin, Alexander and Gharibyan, Hrant and Grover, Tarun and Tai, M. Eric and Rispoli, Matthew and Schittko, Robert and Preiss, Philipp M. and Kaufman, Adam M. and Greiner, Markus and Pichler, Hannes and Hayden, Patrick},
  year = 2019,
  month = jul,
  journal = {Physical Review X},
  volume = {9},
  number = {3},
  pages = {031013},
  publisher = {American Physical Society},
  doi = {10.1103/PhysRevX.9.031013},
  urldate = {2026-05-27}
}

@misc{cui2025unitary,
  title = {Unitary Designs in Nearly Optimal Depth},
  author = {Cui, Laura and Schuster, Thomas and Brandao, Fernando and Huang, Hsin-Yuan},
  year = 2025,
  month = jul,
  number = {arXiv:2507.06216},
  eprint = {2507.06216},
  primaryclass = {quant-ph},
  publisher = {arXiv},
  doi = {10.48550/arXiv.2507.06216},
  urldate = {2026-06-12},
  archiveprefix = {arXiv}
}

@article{daley2012measuring,
  title = {Measuring {{Entanglement Growth}} in {{Quench Dynamics}} of {{Bosons}} in an {{Optical Lattice}}},
  author = {Daley, A. J. and Pichler, H. and Schachenmayer, J. and Zoller, P.},
  year = 2012,
  month = jul,
  journal = {Physical Review Letters},
  volume = {109},
  number = {2},
  pages = {020505},
  issn = {0031-9007, 1079-7114},
  doi = {10.1103/PhysRevLett.109.020505},
  urldate = {2026-05-20},
  copyright = {http://link.aps.org/licenses/aps-default-license},
  langid = {english}
}

@article{ekert2002direct,
  title = {Direct {{Estimations}} of {{Linear}} and {{Nonlinear Functionals}} of a {{Quantum State}}},
  author = {Ekert, Artur K. and Alves, Carolina Moura and Oi, Daniel K. L. and Horodecki, Micha{\l} and Horodecki, Pawe{\l} and Kwek, L. C.},
  year = 2002,
  month = may,
  journal = {Physical Review Letters},
  volume = {88},
  number = {21},
  pages = {217901},
  publisher = {American Physical Society},
  doi = {10.1103/PhysRevLett.88.217901},
  urldate = {2026-06-15}
}

@article{elben2018renyi,
  title = {R\'enyi {{Entropies}} from {{Random Quenches}} in {{Atomic Hubbard}} and {{Spin Models}}},
  author = {Elben, A. and Vermersch, B. and Dalmonte, M. and Cirac, J. I. and Zoller, P.},
  year = 2018,
  month = feb,
  journal = {Physical Review Letters},
  volume = {120},
  number = {5},
  pages = {050406},
  issn = {0031-9007, 1079-7114},
  doi = {10.1103/PhysRevLett.120.050406},
  urldate = {2026-05-20},
  langid = {english}
}

@article{elben2020mixedstate,
  title = {Mixed-{{State Entanglement}} from {{Local Randomized Measurements}}},
  author = {Elben, Andreas and Kueng, Richard and Huang, Hsin-Yuan (Robert) and Van Bijnen, Rick and Kokail, Christian and Dalmonte, Marcello and Calabrese, Pasquale and Kraus, Barbara and Preskill, John and Zoller, Peter and Vermersch, Beno{\^i}t},
  year = 2020,
  month = nov,
  journal = {Physical Review Letters},
  volume = {125},
  number = {20},
  pages = {200501},
  issn = {0031-9007, 1079-7114},
  doi = {10.1103/PhysRevLett.125.200501},
  urldate = {2026-05-27},
  langid = {english}
}

@article{elben2023randomized,
  title = {The Randomized Measurement Toolbox},
  author = {Elben, Andreas and Flammia, Steven T. and Huang, Hsin-Yuan and Kueng, Richard and Preskill, John and Vermersch, Beno{\^i}t and Zoller, Peter},
  year = 2023,
  month = jan,
  journal = {Nature Reviews Physics},
  volume = {5},
  number = {1},
  pages = {9--24},
  publisher = {Nature Publishing Group},
  issn = {2522-5820},
  doi = {10.1038/s42254-022-00535-2},
  urldate = {2026-05-27},
  copyright = {2022 Springer Nature Limited},
  langid = {english}
}

@article{fernandes2024unitaryinvariant,
  title = {Unitary-{{Invariant Witnesses}} of {{Quantum Imaginarity}}},
  author = {Fernandes, Carlos and Wagner, Rafael and Novo, Leonardo and Galv{\~a}o, Ernesto F.},
  year = 2024,
  month = nov,
  journal = {Physical Review Letters},
  volume = {133},
  number = {19},
  pages = {190201},
  issn = {0031-9007, 1079-7114},
  doi = {10.1103/PhysRevLett.133.190201},
  urldate = {2026-06-11},
  langid = {english}
}

@article{foulds2021controlled,
  title = {The Controlled {{SWAP}} Test for Determining Quantum Entanglement},
  author = {Foulds, Steph and Kendon, Viv and Spiller, Tim},
  year = 2021,
  month = apr,
  journal = {Quantum Science and Technology},
  volume = {6},
  number = {3},
  pages = {035002},
  publisher = {IOP Publishing},
  issn = {2058-9565},
  doi = {10.1088/2058-9565/abe458},
  urldate = {2026-05-27},
  langid = {english}
}

@article{haah2017sampleoptimal,
  title = {Sample-{{Optimal Tomography}} of {{Quantum States}}},
  author = {Haah, Jeongwan and Harrow, Aram W. and Ji, Zhengfeng and Wu, Xiaodi and Yu, Nengkun},
  year = 2017,
  month = sep,
  journal = {IEEE Transactions on Information Theory},
  volume = {63},
  number = {9},
  pages = {5628--5641},
  issn = {1557-9654},
  doi = {10.1109/TIT.2017.2719044},
  urldate = {2026-06-12}
}

@article{haferkamp2022random,
  title = {Random {{Quantum Circuits Are Approximate Unitary}} $t$-{{Designs}} in {{Depth}} ${O}(Nt^{5+o(1)})$},
  author = {Haferkamp, Jonas},
  year = 2022,
  month = sep,
  journal = {Quantum},
  volume = {6},
  pages = {795},
  publisher = {Verein zur F\"orderung des Open Access Publizierens in den Quantenwissenschaften},
  doi = {10.22331/q-2022-09-08-795},
  urldate = {2026-06-12},
  langid = {british}
}

@article{hill1997entanglement,
  title = {Entanglement of a {{Pair}} of {{Quantum Bits}}},
  author = {Hill, Sam A. and Wootters, William K.},
  year = 1997,
  month = jun,
  journal = {Physical Review Letters},
  volume = {78},
  number = {26},
  pages = {5022--5025},
  publisher = {American Physical Society},
  doi = {10.1103/PhysRevLett.78.5022},
  urldate = {2026-05-26}
}

@article{huang2020predicting,
  title = {Predicting Many Properties of a Quantum System from Very Few Measurements},
  author = {Huang, Hsin-Yuan and Kueng, Richard and Preskill, John},
  year = 2020,
  month = oct,
  journal = {Nature Physics},
  volume = {16},
  number = {10},
  pages = {1050--1057},
  publisher = {Nature Publishing Group},
  issn = {1745-2481},
  doi = {10.1038/s41567-020-0932-7},
  urldate = {2026-06-15},
  copyright = {2020 The Author(s), under exclusive licence to Springer Nature Limited},
  langid = {english}
}

@article{huggins2021virtual,
  title = {Virtual {{Distillation}} for {{Quantum Error Mitigation}}},
  author = {Huggins, William J. and McArdle, Sam and O'Brien, Thomas E. and Lee, Joonho and Rubin, Nicholas C. and Boixo, Sergio and Whaley, K. Birgitta and Babbush, Ryan and McClean, Jarrod R.},
  year = 2021,
  month = nov,
  journal = {Physical Review X},
  volume = {11},
  number = {4},
  pages = {041036},
  publisher = {American Physical Society},
  doi = {10.1103/PhysRevX.11.041036},
  urldate = {2026-05-27}
}

@article{johri2017entanglement,
  title = {Entanglement Spectroscopy on a Quantum Computer},
  author = {Johri, Sonika and Steiger, Damian S. and Troyer, Matthias},
  year = 2017,
  month = nov,
  journal = {Physical Review B},
  volume = {96},
  number = {19},
  pages = {195136},
  publisher = {American Physical Society},
  doi = {10.1103/PhysRevB.96.195136},
  urldate = {2026-05-27}
}

@article{kada2008efficiency,
  title = {The Efficiency of Quantum Identity Testing of Multiple States},
  author = {Kada, Masaru and Nishimura, Harumichi and Yamakami, Tomoyuki},
  year = 2008,
  month = sep,
  journal = {Journal of Physics A: Mathematical and Theoretical},
  volume = {41},
  number = {39},
  pages = {395309},
  issn = {1751-8121},
  doi = {10.1088/1751-8113/41/39/395309},
  urldate = {2026-05-27},
  langid = {english}
}

@article{knorzer2026distributed,
  title = {Distributed Quantum Information Processing: A Review of Recent Progress},
  shorttitle = {Distributed Quantum Information Processing},
  author = {Kn{\"o}rzer, Johannes and Liu, Xiaoyu and Schiffer, Benjamin and Tura Brugu{\'e}s, Jordi},
  year = 2026,
  journal = {Reports on Progress in Physics},
  issn = {0034-4885},
  doi = {10.1088/1361-6633/ae74e0},
  urldate = {2026-06-12},
  langid = {english}
}

@article{koczor2021exponential,
  title = {Exponential {{Error Suppression}} for {{Near-Term Quantum Devices}}},
  author = {Koczor, B{\'a}lint},
  year = 2021,
  month = sep,
  journal = {Physical Review X},
  volume = {11},
  number = {3},
  pages = {031057},
  publisher = {American Physical Society},
  doi = {10.1103/PhysRevX.11.031057},
  urldate = {2026-05-27}
}

@misc{kostenberger2021weingarten,
  title = {Weingarten {{Calculus}}},
  author = {K{\"o}stenberger, Georg},
  year = 2021,
  month = feb,
  number = {arXiv:2101.00921},
  eprint = {2101.00921},
  primaryclass = {math.PR},
  publisher = {arXiv},
  doi = {10.48550/arXiv.2101.00921},
  urldate = {2026-06-16},
  archiveprefix = {arXiv}
}

@misc{laborde2024quantum,
  title = {Quantum {{Algorithms}} for {{Realizing Symmetric}}, {{Asymmetric}}, and {{Antisymmetric Projectors}}},
  author = {LaBorde, Margarite L. and Rethinasamy, Soorya and Wilde, Mark M.},
  year = 2024,
  month = jul,
  number = {arXiv:2407.17563},
  eprint = {2407.17563},
  primaryclass = {quant-ph},
  publisher = {arXiv},
  doi = {10.48550/arXiv.2407.17563},
  urldate = {2026-05-27},
  archiveprefix = {arXiv}
}

@misc{li2026quantum,
  title = {Quantum {{Nonlinear Properties}} from a {{Single Measurement Setting}}},
  author = {Li, Zihao and Chen, Datong and Qin, Dayue and Yang, Yuxiang and Zhou, You},
  year = 2026,
  month = may,
  number = {arXiv:2605.09958},
  eprint = {2605.09958},
  primaryclass = {quant-ph},
  publisher = {arXiv},
  doi = {10.48550/arXiv.2605.09958},
  urldate = {2026-06-16},
  archiveprefix = {arXiv}
}

@article{linke2018measuring,
  title = {Measuring the {{R\'enyi}} Entropy of a Two-Site {{Fermi-Hubbard}} Model on a Trapped Ion Quantum Computer},
  author = {Linke, N. M. and Johri, S. and Figgatt, C. and Landsman, K. A. and Matsuura, A. Y. and Monroe, C.},
  year = 2018,
  month = nov,
  journal = {Physical Review A},
  volume = {98},
  number = {5},
  pages = {052334},
  issn = {2469-9926, 2469-9934},
  doi = {10.1103/PhysRevA.98.052334},
  urldate = {2026-05-20},
  langid = {english}
}

@article{liu2025generalized,
  title = {Generalized Concentratable Entanglement via Parallelized Permutation Tests},
  author = {Liu, Xiaoyu and Kn{\"o}rzer, Johannes and Wang, Zherui Jerry and Tura, Jordi},
  year = 2025,
  month = jul,
  journal = {Physical Review Research},
  volume = {7},
  number = {3},
  pages = {L032022},
  publisher = {American Physical Society},
  doi = {10.1103/jtlj-qs3y},
  urldate = {2026-05-26}
}

@misc{liu2025measuring,
  title = {Measuring Multipartite Entanglement Efficiently by Testing Symmetries},
  author = {Liu, Xiaoyu and Tura, Jordi and Rico, Albert},
  year = 2025,
  month = nov,
  number = {arXiv:2511.07537},
  eprint = {2511.07537},
  primaryclass = {quant-ph},
  publisher = {arXiv},
  doi = {10.48550/arXiv.2511.07537},
  urldate = {2026-06-15},
  archiveprefix = {arXiv}
}

@book{macdonald1995symmetric,
  title = {Symmetric {{Functions}} and {{Hall Polynomials}}},
  author = {Macdonald, I G},
  year = 1995,
  month = mar,
  publisher = {Oxford University Press},
  doi = {10.1093/oso/9780198534891.001.0001},
  urldate = {2026-06-15},
  isbn = {978-0-19-853489-1}
}

@article{meyer2002global,
  title = {Global Entanglement in Multiparticle Systems},
  author = {Meyer, David A. and Wallach, Nolan R.},
  year = 2002,
  month = sep,
  journal = {Journal of Mathematical Physics},
  volume = {43},
  number = {9},
  pages = {4273--4278},
  issn = {0022-2488},
  doi = {10.1063/1.1497700},
  urldate = {2026-05-26}
}

@article{muller-lennert2013quantum,
  title = {On Quantum {{R\'enyi}} Entropies: {{A}} New Generalization and Some Properties},
  shorttitle = {On Quantum {{R\'enyi}} Entropies},
  author = {{M{\"u}ller-Lennert}, Martin and Dupuis, Fr{\'e}d{\'e}ric and Szehr, Oleg and Fehr, Serge and Tomamichel, Marco},
  year = 2013,
  month = dec,
  journal = {Journal of Mathematical Physics},
  volume = {54},
  number = {12},
  pages = {122203},
  issn = {0022-2488, 1089-7658},
  doi = {10.1063/1.4838856},
  urldate = {2026-05-20},
  langid = {english}
}

@article{quek2024multivariate,
  title = {Multivariate Trace Estimation in Constant Quantum Depth},
  author = {Quek, Yihui and Kaur, Eneet and Wilde, Mark M.},
  year = 2024,
  month = jan,
  journal = {Quantum},
  volume = {8},
  pages = {1220},
  issn = {2521-327X},
  doi = {10.22331/q-2024-01-10-1220},
  urldate = {2026-06-11},
  langid = {english}
}

@incollection{renyi1961measures,
  title = {On {{Measures}} of {{Entropy}} and {{Information}}},
  booktitle = {Proceedings of the {{Fourth Berkeley Symposium}} on {{Mathematical Statistics}} and {{Probability}}, {{Volume}} 1: {{Contributions}} to the {{Theory}} of {{Statistics}}},
  author = {R{\'e}nyi, Alfr{\'e}d},
  year = 1961,
  month = jan,
  volume = {4.1},
  pages = {547--562},
  publisher = {University of California Press},
  urldate = {2026-06-15},
  langid = {english}
}

@article{schuster2025random,
  title = {Random Unitaries in Extremely Low Depth},
  author = {Schuster, Thomas and Haferkamp, Jonas and Huang, Hsin-Yuan},
  year = 2025,
  month = jul,
  journal = {Science},
  volume = {389},
  number = {6755},
  pages = {92--96},
  publisher = {American Association for the Advancement of Science},
  doi = {10.1126/science.adv8590},
  urldate = {2026-06-12}
}

@article{shin2025resourceefficient,
  title = {Resource-Efficient Algorithm for Estimating the Trace of Quantum State Powers},
  author = {Shin, Myeongjin and Lee, Junseo and Lee, Seungwoo and Jeong, Kabgyun},
  year = 2025,
  month = aug,
  journal = {Quantum},
  volume = {9},
  pages = {1832},
  issn = {2521-327X},
  doi = {10.22331/q-2025-08-27-1832},
  urldate = {2026-05-20},
  langid = {english}
}

@article{subramanian2021quantum,
  title = {Quantum Algorithm for Estimating {$\alpha$}-{{Renyi}} Entropies of Quantum States},
  author = {Subramanian, Sathyawageeswar and Hsieh, Min-Hsiu},
  year = 2021,
  month = aug,
  journal = {Physical Review A},
  volume = {104},
  number = {2},
  pages = {022428},
  issn = {2469-9926, 2469-9934},
  doi = {10.1103/PhysRevA.104.022428},
  urldate = {2026-05-20},
  langid = {english}
}

@article{vermersch2018unitary,
  title = {Unitary $n$-Designs via Random Quenches in Atomic Hubbard and Spin Models},
  author = {Vermersch, Beno{\^\i}t and Elben, Andreas and Dalmonte, Marcello and Cirac, Juan Ignacio and Zoller, Peter},
  year = 2018,
  journal = {Physical Review A},
  volume = {97},
  pages = {023604}
}

@article{wang2023quantum,
  title = {Quantum {{Algorithms}} for {{Estimating Quantum Entropies}}},
  author = {Wang, Youle and Zhao, Benchi and Wang, Xin},
  year = 2023,
  month = apr,
  journal = {Physical Review Applied},
  volume = {19},
  number = {4},
  pages = {044041},
  issn = {2331-7019},
  doi = {10.1103/PhysRevApplied.19.044041},
  urldate = {2026-05-20},
  langid = {english}
}

@article{wong2001potential,
  title = {Potential Multiparticle Entanglement Measure},
  author = {Wong, Alexander and Christensen, Nelson},
  year = 2001,
  month = mar,
  journal = {Physical Review A},
  volume = {63},
  number = {4},
  pages = {044301},
  issn = {1050-2947, 1094-1622},
  doi = {10.1103/PhysRevA.63.044301},
  urldate = {2026-05-26},
  copyright = {http://link.aps.org/licenses/aps-default-license},
  langid = {english}
}

@article{yang2021parametrized,
  title = {Parametrized Entanglement Monotone},
  author = {Yang, Xue and Luo, Ming-Xing and Yang, Yan-Han and Fei, Shao-Ming},
  year = 2021,
  month = may,
  journal = {Physical Review A},
  volume = {103},
  number = {5},
  pages = {052423},
  publisher = {American Physical Society},
  doi = {10.1103/PhysRevA.103.052423},
  urldate = {2026-05-26}
}

@misc{ye2025exponential,
  title = {Exponential {{Advantage}} from {{One More Replica}} in {{Estimating Nonlinear Properties}} of {{Quantum States}}},
  author = {Ye, Qi and Liu, Zhenhuan and Deng, Dong-Ling},
  year = 2025,
  month = sep,
  number = {arXiv:2509.24000},
  eprint = {2509.24000},
  primaryclass = {quant-ph},
  publisher = {arXiv},
  doi = {10.48550/arXiv.2509.24000},
  urldate = {2026-05-27},
  archiveprefix = {arXiv}
}

\onecolumngrid
\section*{End Matter}
\twocolumngrid
\textit{Example of Bargmann invariant}.\textemdash
Bargmann invariants are quantities of the form \(\Delta=\tr(\rho_1\cdots\rho_K)\).
For \(K\geqslant 3\), different cyclic orderings of the density matrices are generally coupled through \cref{eq:general_projected_multivariate_main}.
The simplest nontrivial cases occur for \(K=3\), which we focus on, in the following.
Because of cyclicity of the trace, the only two inequivalent cyclic orderings yield the two third-order Bargmann invariants
\begin{equation}\label{eq:k3-bargmann-invariants}
    \Delta_{123}=\tr(\rho_1 \rho_2 \rho_3), \qquad \Delta_{132} = \tr(\rho_1 \rho_3 \rho_2).
\end{equation}

Applying Theorem~1 to the case \(K=3\) yields
\begin{equation}\label{eq:bargmann-forward}
    \bar\sigma_3^{(d,m)}(\rho_1,\rho_2,\rho_3) =
    \gamma_e\,\Delta_{123} + \gamma_c\,\Delta_{132}
    + \gamma_t\,\Xi+\gamma_c,
\end{equation}
with
\begin{equation}
    \Xi = \tr(\rho_2\rho_3) + \tr(\rho_1\rho_3) + \tr(\rho_1\rho_2),
\end{equation}
where \(\gamma_e\), \(\gamma_c\) and \(\gamma_t\) denote the coefficients associated with the identity permutation, three-cycles and transpositions, respectively, all of which depend on $d$ and $m$.

Similarly,
\begin{equation}\label{eq:bargmann-backward}
    \bar\sigma_3^{(d,m)}(\rho_1,\rho_3,\rho_2) =
    \gamma_e\,\Delta_{132} + \gamma_c\,\Delta_{123} + \gamma_t\,\Xi+\gamma_c.
\end{equation}
Thus the two cyclic orderings are coupled through the projected observables in Eqs.~\eqref{eq:bargmann-forward} and \eqref{eq:bargmann-backward}, which can be expressed as
\begin{equation}
    \mathbf{R} \equiv
    \begin{pmatrix}
        \bar\sigma_3^{(d,m)}(\rho_1,\rho_2,\rho_3)-\gamma_t \Xi-\gamma_c \\
        \bar\sigma_3^{(d,m)}(\rho_1,\rho_3,\rho_2)-\gamma_t \Xi-\gamma_c
    \end{pmatrix} =
    \begin{pmatrix}
        \gamma_e & \gamma_c \\
        \gamma_c & \gamma_e
    \end{pmatrix}
    \begin{pmatrix}
        \Delta_{123}\\
        \Delta_{132}
    \end{pmatrix}.
\end{equation}
Provided \(\gamma_e^2-\gamma_c^2\neq 0\), the two distinct multivariate traces in \cref{eq:k3-bargmann-invariants} are thus given by
\begin{equation}\label{eq:bargmann-reconstruction}
    \begin{pmatrix}
        \Delta_{123}\\
        \Delta_{132}
    \end{pmatrix} =
    \frac{1}{\gamma_e^2-\gamma_c^2}
    \begin{pmatrix}
        \gamma_e & - \gamma_c\\
        -\gamma_c & \gamma_e
    \end{pmatrix}
    \mathbf{R}.
\end{equation}
Since the \(\Delta_{123}\) and \(\Delta_{132}\) are generally complex, both their real and imaginary parts must be estimated independently.
One may reconstruct them from the expectation values of the Hermitian operators \((V_{3}+V_{3}^\dagger)/2\) and \((V_{3}-V_{3}^\dagger)/(2i)\).
Operationally, this corresponds to combining measurements from generalized swap tests in different bases.

As a concrete example, we consider the third-order Bargmann invariant
\begin{equation}
\Delta_{123}(\phi)
=
\langle\psi_1|\psi_2\rangle
\langle\psi_2|\psi_3(\phi)\rangle
\langle\psi_3(\phi)|\psi_1\rangle
\end{equation}
for three four-qubit product states
\begin{align}
|\psi_1\rangle
&=
|0\rangle^{\otimes n},\\
|\psi_2\rangle
&=
\left(
\cos\frac{\theta}{2}\,|0\rangle
+
\sin\frac{\theta}{2}\,|1\rangle
\right)^{\otimes n},\\
|\psi_3(\phi)\rangle
&=
\left(
\cos\frac{\theta}{2}\,|0\rangle
+
e^{i\phi}\sin\frac{\theta}{2}\,|1\rangle
\right)^{\otimes n},
\end{align}
with \(n=4\), \(\theta=\pi/2\), and \(\phi\in[0,2\pi]\).

\begin{figure}[t]
    \centering
    \includegraphics[width=0.99\linewidth]{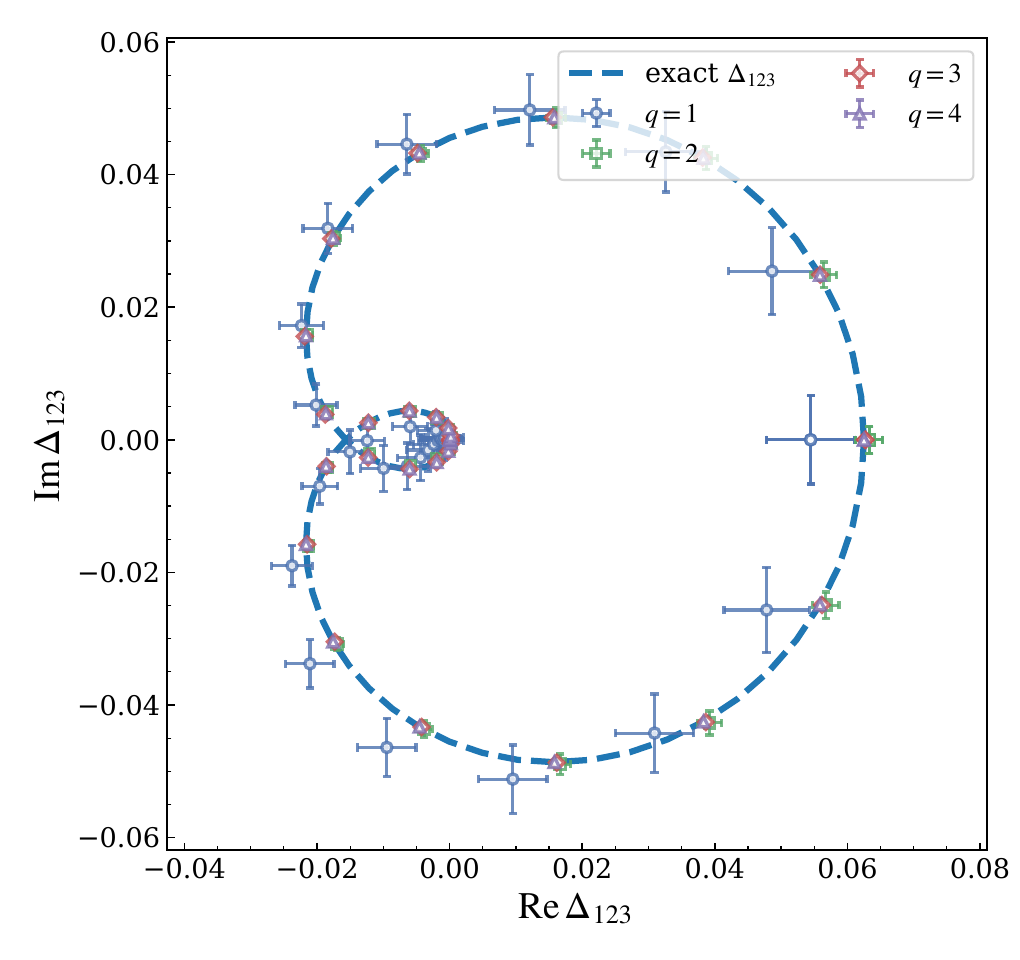}
    \caption{Reconstruction of \(\Delta_{123}\) in the complex plane.
    The dashed curve shows the exact values, while symbols denote estimates obtained from random projections for different numbers \(q\) of remaining qubits.
    Error bars indicate one standard error of the mean obtained from \(50\) independent batches.
    Increasing \(q\) improves the accuracy and reduces statistical fluctuations, with the estimates approaching the exact trajectory.}
    \label{fig:bargmann-K3-example}
\end{figure}

Figure~\ref{fig:bargmann-K3-example} shows the resulting trajectory of
\(\Delta_{123}\) in the complex plane.
The dashed curve corresponds to the exact values, while the markers denote estimates obtained from the random projection protocol for different projection dimensions \(m=2^q\), where \(q\) is the number of retained qubits.

For each value of \(q\), the three pairwise overlaps entering
\(\Xi\)
are estimated independently using the two-copy reconstruction formula, and combined with the estimate of the projected three-copy quantity according to Eq.~\eqref{eq:bargmann-reconstruction}.
We use the same number of $10^4$ protocol executions for each of the estimators.
Since the pairwise overlap estimators consume two copies per execution and the three-copy estimator consumes three copies per execution, the total number of consumed state copies is \(N=5\times10^4\).
Statistical uncertainties are estimated from 50 independent batches and are shown as one standard error of the mean.
As expected, increasing the projection dimension improves the accuracy and reduces statistical fluctuations.

\textit{Partial transpose}.\textemdash
Let \(\rho_{AB}\)
be a bipartite state on \(\mathcal H_A\otimes\mathcal H_B\), and denote
by \(T_B\) the partial transpose with respect to subsystem \(B\).
We define
\begin{equation}
\mu_K
=
\tr\!\left[(\rho_{AB}^{T_B})^K\right].
\end{equation}

Let \(V_K^{(A)}\) and \(V_K^{(B)}\)
denote the cyclic permutation operators acting on the \(A\) and
\(B\) subsystems of \(K\) copies.
To derive a permutation representation
of \(\mu_K\), we write matrix elements in a product basis as
\(
\langle a,b|\rho_{AB}|a',b'\rangle .
\)
The partial transpose on subsystem \(B\) acts as
\begin{equation}
\langle a,b|\rho_{AB}^{T_B}|a',b'\rangle
=
\langle a,b'|\rho_{AB}|a',b\rangle .
\end{equation}
Expanding the trace of \((\rho_{AB}^{T_B})^K\) yields
\begin{align}
\mu_K
&=
\sum_{\substack{a_1,\cdots,a_K\\ b_1,\cdots,b_K}}
\langle a_1,b_1|\rho_{AB}^{T_B}|a_2,b_2\rangle
\cdots
\langle a_K,b_K|\rho_{AB}^{T_B}|a_1,b_1\rangle
\nonumber\\
&=
\sum_{\substack{a_1,\cdots,a_K\\ b_1,\cdots,b_K}}
\langle a_1,b_2|\rho_{AB}|a_2,b_1\rangle
\cdots
\langle a_K,b_1|\rho_{AB}|a_1,b_K\rangle .
\end{align}
Thus the \(A\)-indices are contracted according to
the cyclic permutation \(V_K^{(A)}\), while the \(B\)-indices are contracted in the
opposite cyclic order according to \([V_K^{(B)}]^{-1}\).
Therefore,
\begin{equation}
\mu_K
=
\tr\!\left[
\rho_{AB}^{\otimes K}
\left(
V_{K}^{(A)}\otimes [V_K^{(B)}]^{-1}
\right)
\right].
\label{eq:pt-replica-identity}
\end{equation}

For partial-transpose moments, the relevant observables are therefore
subsystem permutations rather than global cyclic permutations.
We define the subsystem-permutation invariants
\begin{equation}
I_{r,s}(\rho)
=
\tr\!\left[
\rho_{AB}^{\otimes K}
\left(
V^{(A)}_{r}\otimes V^{(B)}_{s}
\right)
\right],
\qquad
r,s\in S_K,
\label{eq:subsystem-invariants}
\end{equation}
where \(V_r^{(A)}\) and \(V_s^{(B)}\) denote the permutation operators corresponding to \(r,s\in S_K\) acting on subsystems \(A\) and \(B\), respectively.
The desired partial-transpose moment corresponds to the choice
\(r=\pi, s=\pi^{-1}\) with \(\pi=(1\,2\,\cdots\,K),\)
that is,
\begin{equation}
\mu_K
=
I_{\pi,\pi^{-1}}(\rho).
\end{equation}

We now derive the corresponding projected observables.
Consider local random projections
\begin{equation}
P^{(A)}_{U}=U P^{(A)} U^\dagger,
\qquad
P^{(B)}_{W}=W P^{(B)} W^\dagger,
\end{equation}
with ranks \(m_A\) and \(m_B\), respectively, and define the projected
state
\begin{equation}
\sigma_{U,W}
=
(P^{(A)}_{U}\otimes P^{(B)}_{W})
\rho_{AB}
(P^{(A)}_{U}\otimes P^{(B)}_{W}) .
\end{equation}
For \(r,s\in S_K\), let
\begin{equation}
M_{r,s}(U,W)
=
\tr\!\left[
\sigma_{U,W}^{\otimes K}
\left(
V^{(A)}_{r}\otimes V^{(B)}_{s}
\right)
\right].
\end{equation}
Using \((P^{(A)}_{U})^2=P^{(A)}_{U}\), \((P^{(B)}_{W})^2=P^{(B)}_{W}\), and cyclicity of the
trace, this becomes
\begin{align}
&M_{r,s}(U,W)
=\nonumber\\
&\tr\!\left[
\rho_{AB}^{\otimes K}
\left(
(P^{(A)}_{U})^{\otimes K}
\otimes
(P^{(B)}_{W})^{\otimes K}
\right)
\left(
V^{(A)}_{r}\otimes V^{(B)}_{s}
\right)
\right].
\end{align}
Averaging independently over the two Haar-random unitaries gives, for \(\overline M_{r,s} \equiv \mathbb E_{U,W}\!\left[M_{r,s}(U,W)\right]\),
\begin{equation}
\overline M_{r,s} =
\tr\!\left[
\rho_{AB}^{\otimes K}
\left(
\Omega_K^{(A)}
\otimes
\Omega_K^{(B)}
\right)
\left(
V^{(A)}_{r}\otimes V^{(B)}_{s}
\right)
\right].
\end{equation}
where \(\Omega_K^{(A)} = \mathbb E_U[(P^{(A)}_{U})^{\otimes K}]\) and \(\Omega_K^{(B)} = \mathbb E_W[(P^{(B)}_{W})^{\otimes K}]\).

Applying the projector-twirling identity separately on the two
subsystems \(X=A, B\) yields
\begin{equation}
\Omega_K^{(X)}
=
\sum_{\alpha\in S_K}
\gamma_\alpha(d_X,m_X)
V^{(X)}_{\alpha}.
\end{equation}
Substituting these expansions and using
\(V^{(X)}_\alpha V^{(X)}_r = V^{(X)}_{\alpha r}\), we obtain
\begin{equation}
\overline M_{r,s}
=
\sum_{\alpha,\beta\in S_K}
\gamma_\alpha(d_A,m_A)
\gamma_\beta(d_B,m_B)
I_{\alpha r,\beta s}(\rho).
\label{eq:pt-projected-reconstruction}
\end{equation}

Collecting the
projected observables into a vector \(\mathbf M\) and the invariants into
a vector \(\mathbf I\), Eq.~\eqref{eq:pt-projected-reconstruction}
defines a linear system
\begin{equation}
\mathbf M
=
C\,\mathbf I,
\end{equation}
where the matrix \(C\) depends only on the dimensions
\(d_A,d_B\) and projection ranks \(m_A,m_B\).
Whenever \(C\) is invertible,
\begin{equation}
\mathbf I
=
C^{-1}\mathbf M .
\end{equation}
In particular, the third partial-transpose moment is obtained from
\begin{equation}
\mu_3
=
\tr\!\left[(\rho_{AB}^{T_B})^3\right]
=
I_{(123),(132)}
=
\left(C^{-1}\mathbf M\right)_{(123),(132)} .
\end{equation}

Compared to ordinary state moments, the required coherent quantum
processing remains unchanged.
Estimating \(\mu_K\) still requires \(K\)-copy measurements after projection onto lower-dimensional subspaces.
The difference is that the measurement now probes
subsystem permutations \(V^{(A)}_{r}\otimes V^{(B)}_{s}\) rather than a single
global permutation operator.
Consequently, the required coherent quantum resources
are still determined by the copy number \(K\), while the reconstruction
requires solving for a larger set of subsystem-permutation invariants.

\onecolumngrid
\newpage

\setcounter{section}{0}
\renewcommand{\thesection}{S\arabic{section}}
\renewcommand{\thesubsection}{S\arabic{section}.\arabic{subsection}}

\begin{center}
\textbf{\large Supplemental Materials}
\end{center}

\section{Preliminaries and notation \label{sec:preliminaries}}

In this appendix, we review the generalized swap test and the representation-theoretic framework underlying the projected moment identities presented in the main text.
The central objects appearing throughout the work are permutation operators acting on tensor-product spaces and their behavior under Haar averaging.
These structures naturally lead to Schur--Weyl duality and Weingarten calculus.

\subsection{Permutation operators and trace invariants}

Let \(\mathcal H \cong \mathbb C^d\) and consider the tensor-product space \(\mathcal H^{\otimes K}\).
The symmetric group \(S_K\) acts naturally on \(\mathcal H^{\otimes K}\) by permuting the \(K\) tensor factors.
For each permutation \(\pi \in S_K\), we denote by \(V_\pi : \mathcal H^{\otimes K} \to \mathcal H^{\otimes K}\) the corresponding permutation operator,
\begin{equation}
V_\pi
\bigl(
|i_1\rangle \otimes \cdots \otimes |i_K\rangle
\bigr)
=
|i_{\pi(1)}\rangle
\otimes \cdots \otimes
|i_{\pi(K)}\rangle .
\end{equation}
Throughout, products of permutations are composed in the convention compatible with
\(V_\tau V_\pi=V_{\tau\pi}\).
A particularly important role is played by cyclic permutations.
We write \(V_K \equiv V_{(1\,2\,\cdots\,K)}\) for the cyclic permutation operator on \(K\) copies, defined by
\begin{equation}
V_K
\bigl(
|i_1\rangle\otimes\cdots\otimes|i_K\rangle
\bigr)
=
|i_2\rangle\otimes\cdots\otimes|i_K\rangle\otimes|i_1\rangle.
\end{equation}

The generalized swap trick states that
\begin{equation}
\tr(\rho^K)
=
\tr\!\left(
V_K \rho^{\otimes K}
\right).
\label{eq:app-swap}
\end{equation}
To see this explicitly, let
\(
\rho=\sum_{i,j}\rho_{ij}|i\rangle\langle j|.
\)
Then
\begin{align}
\operatorname{tr}\!\left(V_K\rho^{\otimes K}\right)
&=
\sum_{i_1,\ldots,i_K}
\bra{i_1,\ldots,i_K}
V_K\rho^{\otimes K}
\ket{i_1,\ldots,i_K} \notag\\
&=
\sum_{i_1,\ldots,i_K}
\bra{i_K,i_1,\ldots,i_{K-1}}
\rho^{\otimes K}
\ket{i_1,\ldots,i_K} \notag\\
&=
\sum_{i_1,\ldots,i_K}
\rho_{i_K i_1}
\rho_{i_1 i_2}
\cdots
\rho_{i_{K-1}i_K}
=
\operatorname{tr}(\rho^K).
\end{align}

Thus the cyclic permutation contracts the output index of each copy with the input index of the next copy, producing the matrix product \(\rho^K\) under the trace.
More generally, permutation operators convert multi-copy expectation values into nonlinear trace invariants determined by cycle structure.
If \(\pi \in S_K\) decomposes into disjoint cycles,
\(
\pi = c_1 c_2 \cdots c_r,
\)
then
\begin{equation}
\tr\!\left(
V_\pi \rho^{\otimes K}
\right)
=
\prod_{j=1}^r
\tr\!\left(
\rho^{|c_j|}
\right),
\label{eq:app-cycle-factorization}
\end{equation}
where \(|c_j|\) denotes the length of the cycle \(c_j\).
Consequently, polynomial functions of quantum states are naturally organized by permutation structure in \(S_K\).

\subsection{Generalized swap test}

The generalized swap test estimates \(p_K=\tr(\rho^K)\) by coherently controlling a permutation operator acting on multiple copies of the state.
It can be regarded as a Hadamard test with coherently controlled $V_K$.
For input states $\rho_1,\cdots,\rho_K$, estimating $\tr(\rho_1\cdots\rho_K)$ can be divided into estimating its real and imaginary parts.
For the real part, the two corresponding POVM elements for measuring $\ket{0}$ and $\ket{1}$ on the ancillary qubit are $E_0^{(\mathrm{Re})} = \left( \frac{\mathbb{I}+V_K}{2} \right)^{\dagger}\left( \frac{\mathbb{I}+V_K}{2} \right)$ and $E_1^{(\mathrm{Re})} = \left( \frac{\mathbb{I}-V_K}{2} \right)^{\dagger}\left( \frac{\mathbb{I}-V_K}{2} \right)$.
For the imaginary part, the two corresponding POVM elements for measuring $\ket{0}$ and $\ket{1}$ are $E_0^{(\mathrm{Im})} = \left( \frac{\mathbb{I}-iV_K}{2} \right)^{\dagger}\left( \frac{\mathbb{I}-iV_K}{2} \right)$ and $E_1^{(\mathrm{Im})} = \left( \frac{\mathbb{I}+iV_K}{2} \right)^{\dagger}\left( \frac{\mathbb{I}+iV_K}{2} \right)$.
By estimating the probability of obtaining $\ket{0}$ or $\ket{1}$ in each case, one obtains:
\begin{equation}
    \mathrm{Re}[\tr(\rho_1\cdots\rho_K)] = \tr(E_0^{(\mathrm{Re})} (\rho_1\otimes\cdots\otimes\rho_K)) -  \tr(E_1^{(\mathrm{Re})} (\rho_1\otimes\cdots\otimes\rho_K)),
\end{equation}
and,
\begin{equation}
    \mathrm{Im}[\tr(\rho_1\cdots\rho_K)] = \tr(E_0^{(\mathrm{Im})} (\rho_1\otimes\cdots\otimes\rho_K)) -  \tr(E_1^{(\mathrm{Im})} (\rho_1\otimes\cdots\otimes\rho_K)).
\end{equation}
In particular, when $\rho_1=\cdots=\rho_K=\rho$, the corresponding value is real, and therefore:
\begin{equation}
    p_K = \tr(\rho^K) = \tr(E_0^{(\mathrm{Re})} \rho^{\otimes K}) -  \tr(E_1^{(\mathrm{Re})} \rho^{\otimes K}).
\end{equation}

\subsection{Haar averaging and Schur--Weyl duality}

The random projection protocol introduced in the main text is based on Haar-random projectors
\(
P_U = U P U^\dagger\) with \(U \sim \mu_{\rm H},\)
where \(P\) is a fixed rank-\(m\) projector and \(\mu_{\rm H}\) denotes the Haar measure on the unitary group \(U(d)\).
Projected moments therefore involve Haar averages of operators on tensor-product spaces.
For an operator
\(
O \in \mathcal L(\mathcal H^{\otimes K}),
\)
we define the Haar-twirling channel
\begin{equation}
\Phi_K(O)
=
\int_{U(d)}
U^{\otimes K}
\, O \,
(U^\dagger)^{\otimes K}
\, d\mu_{\rm H}(U).
\end{equation}

A fundamental result in the representation theory of the unitary and symmetric groups is that Haar twirling projects operators onto the commutant of the tensor-power representation \(U^{\otimes K}\), namely the set of operators commuting with \(U^{\otimes K}\) for every \(U\in U(d)\).
Schur--Weyl duality states that this commutant is generated by permutations of the tensor factors, i.e.,
\(
\mathrm{Comm}
\!\left(
\{U^{\otimes K}:U\in U(d)\}
\right)
=
\mathrm{span}
\{
V_\pi:\pi\in S_K
\}.
\)
As a consequence, every Haar-twirled operator can be expanded in the permutation basis,
\begin{equation}
\Phi_K(O)
=
\sum_{\pi\in S_K}
c_\pi(O)\,
V_\pi .
\label{eq:app-permutation-expansion}
\end{equation}
Applying \cref{eq:app-permutation-expansion} to
\(
O=P^{\otimes K}
\)
leads to the permutation expansion underlying Theorem~1.
The coefficients \(c_\pi(O)\) are determined using Weingarten calculus, the necessary basics of which are briefly reviewed next.

\subsection{Weingarten calculus}

To determine the coefficients appearing in the permutation expansion, we use Weingarten calculus.
This provides explicit formulas for Haar averages of tensor powers of unitary matrices and their conjugates.
For operators on \(\mathcal H^{\otimes K}\), the Haar twirling channel admits the expansion
\begin{equation}
\Phi_K(O)
=
\sum_{\alpha,\beta\in S_K}
\mathrm{Wg}_d(\alpha^{-1}\beta)
\,
\tr\!\left(
O V_{\alpha^{-1}}
\right)
V_\beta,
\label{eq:app-weingarten}
\end{equation}
where \(\mathrm{Wg}_d:S_K\to\mathbb R\) denotes the unitary Weingarten function.
It provides the coefficients appearing when Haar averages are expanded in the permutation basis.
For example, when \(K=2\), \(S_2=\{e,(12)\},\) and the Weingarten function evaluates to
\begin{equation}
\mathrm{Wg}_d(e)
=
\frac{1}{d^2-1},
\qquad
\mathrm{Wg}_d((12))
=
-\frac{1}{d(d^2-1)}.
\end{equation}
Substituting these coefficients into Eq.~\eqref{eq:app-weingarten} reproduces the familiar decomposition of second moments into symmetric and antisymmetric sectors.

For the tensor-product projector
\[
O=P^{\otimes K},
\]
the traces appearing in Eq.~\eqref{eq:app-weingarten} simplify considerably.
Indeed, if \(P\) has rank \(m\), then
\begin{equation}
\tr\!\left(
P^{\otimes K}V_\alpha
\right)
=
m^{\mathrm c(\alpha)},
\label{eq:app-projector-trace}
\end{equation}
where $\mathrm{c}(\alpha)$ denotes the number of cycles in $\alpha\in S_K$.
This follows from the same cycle-contraction mechanism underlying the generalized swap trick.
Writing \(A=\sum_{i,j}A_{ij}|i\rangle\langle j|,\)
one finds
\begin{equation}
\tr\!\left(
A^{\otimes K}V_\alpha
\right)
=
\sum_{i_1,\cdots,i_K}
A_{i_1,i_{\alpha(1)}}
\cdots
A_{i_K,i_{\alpha(K)}}.
\end{equation}
The indices contract independently along the cycles of \(\alpha\).
More precisely, if \(\alpha=\omega_1\cdots\omega_r\) is the decomposition of \(\alpha\) into disjoint cycles, then each cycle
\(\omega_j=(a_1\,a_2\,\cdots\,a_\ell)\) contributes a factor
\begin{equation}
\sum_{i_{a_1},\cdots,i_{a_\ell}}
A_{i_{a_1}i_{a_2}}
A_{i_{a_2}i_{a_3}}
\cdots
A_{i_{a_\ell}i_{a_1}}
=
\tr(A^\ell).
\end{equation}
Consequently,
\begin{equation}
\tr\!\left(
A^{\otimes K}V_\alpha
\right)
=
\prod_{\omega\in\mathrm{cyc}(\alpha)}
\tr(A^{|\omega|}),
\end{equation}
where \(\mathrm{cyc}(\alpha)\) denotes the set of cycles of \(\alpha\).
Applying this identity to the rank-\(m\) projector \(P\), we obtain
\begin{equation}
\tr\!\left(
P^{\otimes K}V_\alpha
\right)
=
\prod_{\omega\in\mathrm{cyc}(\alpha)}
\tr(P^{|\omega|}).
\end{equation}
Since \(P\) is a projector, \(P^{|\omega|}=P\) for every cycle \(\omega\).
Therefore each cycle contributes a factor \(\tr(P)=m,\) which proves \cref{eq:app-projector-trace}.

\subsection{Character-theoretic ingredients}

An alternative description of the permutation algebra is obtained through the representation theory of the symmetric group \(S_K\).
The irreducible representations of \(S_K\) are indexed by partitions \(\lambda\vdash K.\)
For example, the partitions of \(K=3\) are \((3), (2,1), (1,1,1).\)
The partition \((2,1)\) corresponds to the Young diagram
\[
\ydiagram{2,1},
\]
with two boxes in the first row and one box in the second row.

\emph{Dimensions.}\textemdash\
The quantity \(f^\lambda\) denotes the dimension of the irreducible representation associated with the partition \(\lambda\).
This dimension is determined by the hook-length formula
\begin{equation}
f^\lambda
=
\frac{K!}
{\prod_{(i,j)\in\lambda} h_{ij}},
\label{eq:app-hook}
\end{equation}
where \(h_{ij}\) denotes the hook length of the box \((i,j)\) in the Young diagram, i.e., the number of boxes directly to its right and below, including the box itself.
For \(\lambda=(2,1)\), the hook lengths are \(3\), \(1\), and \(1\), giving
\(
f^{(2,1)}
=
3!/(3\cdot1\cdot1)
=
2.
\)

\emph{Characters.}\textemdash\
If \(\rho_\lambda:S_K\to GL(V_\lambda)\)
denotes the irreducible representation associated with \(\lambda\), then its character is defined by
\begin{equation}
\chi_\lambda(\pi)
=
\tr\!\bigl(\rho_\lambda(\pi)\bigr).
\label{eq:app-character}
\end{equation}
Characters encode how permutations act inside a given irreducible representation.

For the partition \((2,1)\), the corresponding irreducible representation of \(S_3\) is two-dimensional.
The transposition \((12)\) has character \(\chi^{(2,1)}((12))=0.\)
Indeed, in this representation transpositions act as reflections, whose eigenvalues are \(+1\) and \(-1\), so their trace vanishes.

\emph{Content polynomials.}\textemdash\
Associated with each partition is the content polynomial
\begin{equation}
\alpha_\lambda(x)
=
\prod_{(i,j)\in\lambda}
(x+j-i),
\label{eq:app-content}
\end{equation}
where the product runs over all boxes \((i,j)\) of the Young diagram.

For \(\lambda=(2,1)\), the boxes are located at \((1,1), \ (1,2)\) and \((2,1),\) which produce the factors \(x, \ x+1\) and \( x-1.\)
Hence
\(
\alpha_{(2,1)}(x)
=
x(x+1)(x-1).
\)

These quantities enter the character expansion of the Weingarten function and, ultimately, the coefficient formula of Theorem~1 derived in Appendix~\ref{sec:supp-proof-theorem1}.

\section{Proof of main results \label{sec:supp-proof-theorem1}}

In this section we prove Theorem~1, Corollary~1 and derive the expansion coefficients \(\gamma_\tau(d,m)\) from \cref{eq:coefficients-weingarten}.
The derivation of the main result combines the generalized swap identity
Eq.~\eqref{eq:app-swap}, the Schur--Weyl decomposition
Eq.~\eqref{eq:app-permutation-expansion}, and the Weingarten expansion
Eq.~\eqref{eq:app-weingarten} reviewed in \cref{sec:preliminaries}.

\subsection{Theorem 1: general multivariate traces \label{ssec:thm1-multivariate-traces}}

Let \(\sigma_{i,U}=P_U\rho_i P_U, P_U=UPU^\dagger,\) where \(P\) is a rank-\(m\) projector on \(\mathbb C^d\), and let \(\pi=(1\,2\,\cdots\,K)\in S_K\) denote the cyclic permutation.
Using the generalized swap identity \cref{eq:app-swap},
\begin{equation}\tr(\sigma_{1,U}\cdots \sigma_{K,U}) =
\tr\!\left[(\sigma_{1,U}\otimes\cdots\otimes\sigma_{K,U})V_K\right]  =
\tr\!\left[ (\rho_1\otimes\cdots\otimes\rho_K)P_U^{\otimes K}V_K\right].
\end{equation}

Averaging over the Haar measure gives
\begin{align}
\bar\sigma_K^{(d,m)}(\rho_1,\cdots,\rho_K)
&=
\tr\!\left[
(\rho_1\otimes\cdots\otimes\rho_K)
\Phi_K(P^{\otimes K})
V_K
\right],
\label{eq:supp-general-1}
\end{align}
where
\begin{equation}
\Phi_K(O)=\int_{U(d)}U^{\otimes K}\,O\,(U^\dagger)^{\otimes K}\,dU
\end{equation}
denotes the \(K\)-fold Haar twirling channel.
By Eq.~\eqref{eq:app-permutation-expansion},
\begin{equation}\label{eq:expansion-phiK}
\Phi_K(P^{\otimes K})
=
\sum_{\tau\in S_K}
\gamma_\tau(d,m)\,
V_\tau .
\end{equation}
Substituting this expansion into \cref{eq:supp-general-1} yields
\begin{align}
\bar\sigma_K^{(d,m)}(\rho_1,\cdots,\rho_K)
&=
\sum_{\tau\in S_K}
\gamma_\tau(d,m)
\tr\!\left[
(\rho_1\otimes\cdots\otimes\rho_K)
V_{\tau\pi}
\right].
\label{eq:supp-general-2}
\end{align}

For a permutation \(\tau\pi=C_1\cdots C_{c(\tau\pi)}\) decomposed into disjoint cycles, each cycle \(C_j=(\ell_1\cdots\ell_r)\) contributes the factor \(\tr(\rho_{\ell_1}\cdots\rho_{\ell_r})\), so that
\begin{equation}
\tr\!\left[
(\rho_1\otimes\cdots\otimes\rho_K)
V_{\tau\pi}
\right]
=
\prod_{j=1}^{c(\tau\pi)}
\tr\!\left(
\prod_{\ell\in C_j(\tau\pi)}
\rho_\ell
\right).
\end{equation}

Consequently,
\begin{equation}
\bar\sigma_K^{(d,m)}(\rho_1,\cdots,\rho_K)
=
\sum_{\tau\in S_K}
\gamma_\tau(d,m)
\prod_{j=1}^{c(\tau\pi)}
\tr\!\left(
\prod_{\ell\in C_j(\tau\pi)}
\rho_\ell
\right),
\end{equation}
which proves Theorem~1.
\hfill $\square$

\subsection{Expansion coefficients \label{ssec:expansion-coefficients-derivation}}

It remains to determine the coefficients \(\gamma_\tau(d,m)\).
Using the Weingarten expansion Eq.~\eqref{eq:app-weingarten} with
\(O=P^{\otimes K}\) gives
\begin{equation}
\Phi_K(P^{\otimes K})
=
\sum_{\alpha,\beta\in S_K}
\mathrm{Wg}_d(\alpha^{-1}\beta)
\,
\tr\!\left(
P^{\otimes K}V_{\alpha^{-1}}
\right)
V_\beta.
\end{equation}

Using the projector trace identity \cref{eq:app-projector-trace}, we obtain
\begin{equation}
\Phi_K(P^{\otimes K})
=
\sum_{\alpha,\beta\in S_K}
m^{\mathrm c(\alpha)}
\,
\mathrm{Wg}_d(\alpha^{-1}\beta)
V_\beta .
\end{equation}

A comparison with the permutation expansion in \cref{eq:expansion-phiK} yields
\begin{equation}
\gamma_\tau(d,m)
=
\sum_{\alpha\in S_K}
m^{\mathrm c(\alpha)}
\,
\mathrm{Wg}_d(\alpha^{-1}\tau).
\label{eq:supp-weingarten-coefficients}
\end{equation}

The unitary Weingarten function itself admits the character expansion
\begin{equation}
\mathrm{Wg}_d(\sigma)
=
\frac{1}{K!}
\sum_{\substack{\lambda\vdash K\\
\ell(\lambda)\leqslant d}}
\frac{f^\lambda}
{\alpha_\lambda(d)}
\,
\chi_\lambda(\sigma),
\label{eq:supp-weingarten-character-B}
\end{equation}
where \(\ell(\lambda)\) denotes the number of rows of the partition \(\lambda\).

The restriction \(\ell(\lambda)\leqslant d\) arises from Schur--Weyl duality:
only partitions with at most \(d\) rows occur in the decomposition of
\((\mathbb C^d)^{\otimes K}\).
Equivalently, \(\alpha_\lambda(d)=0\) whenever
\(\ell(\lambda)>d\).
Substituting Eq.~\eqref{eq:supp-weingarten-character-B}
into Eq.~\eqref{eq:supp-weingarten-coefficients} and using the orthogonality relations of irreducible characters yields
\begin{equation}
\gamma_\tau(d,m)
=
\frac1{K!}
\sum_{\substack{\lambda\vdash K,\\
\ell(\lambda)\leqslant d}}
f^\lambda
\frac{\alpha_\lambda(m)}
{\alpha_\lambda(d)}
\chi_\lambda(\tau),
\label{eq:supp-character-coefficients}
\end{equation}
where \(f^\lambda\), \(\chi_\lambda\), and \(\alpha_\lambda\) were introduced in \cref{sec:preliminaries}.
Equation~\eqref{eq:supp-character-coefficients} is the character representation of the coefficients stated in the main text.

\hfill $\square$

\subsection{Corollary 1: identical state copies}

Setting \(\rho_1=\cdots=\rho_K=\rho\)
in Theorem~1, every cycle \(C_j=(\ell_1\cdots\ell_{\nu_j})\) contributes \(\tr(\rho^{\nu_j}),\) where \(\nu_j\) denotes the cycle length.
Hence
\begin{equation}\label{eq:corollary-sm}
\bar\sigma_K^{(d,m)}(\rho)
=
\sum_{\tau\in S_K}
\gamma_\tau(d,m)
\prod_{j=1}^{c(\tau\pi)}
\tr(\rho^{\nu_j}),
\end{equation}
which is precisely the statement of Corollary~1.

\hfill $\square$

\section{Variance of projected moments \label{sec:variance-moments}}

In this section we derive a general expression for the variance of projected moments under Haar-random projections.
The resulting formula is then illustrated for the third projected moment of a noisy five-qubit GHZ state, where it predicts the fluctuations arising solely from the random choice of projection subspace.

\subsection{General variance formula}

Let \(X_K(U):=\tr(\sigma_U^K),\) and \(\sigma_U=P_U\rho P_U.\)
By the generalized swap trick,
\begin{equation}
X_K(U)
=
\tr\!\left(
\rho^{\otimes K}
P_U^{\otimes K}
V_K
\right),\qquad
X_K(U)^2
=
\tr\!\left(
\rho^{\otimes 2K}
P_U^{\otimes 2K}
V_{\Pi_K}
\right),
\label{eq:variance-xk}
\end{equation}
where
\begin{equation}
\Pi_K
=
(1\,2\,\cdots\,K)
(K+1\,K+2\,\cdots\,2K)
\in S_{2K}.
\label{eq:variance-pik}
\end{equation}
Indeed, the product of two traces can be written as a single trace on the tensor-product space of the two independent \(K\)-copy registers.
Averaging over \(U\) gives
\begin{equation}
\mathbb E_U[X_K(U)^2]
=
\tr\!\left[
\rho^{\otimes 2K}
\Phi_{2K}(P^{\otimes 2K})
V_{\Pi_K}
\right].
\label{eq:variance-twirl}
\end{equation}
Using the permutation expansion of \(\Phi_{2K}(P^{\otimes 2K})\) we obtain
\begin{equation}
\mathbb E_U[X_K(U)^2]
=
\sum_{\tau\in S_{2K}}
\gamma^{(2K)}_\tau(d,m)
\tr\!\left(
\rho^{\otimes 2K}
V_{\tau\Pi_K}
\right).
\label{eq:variance-precycle}
\end{equation}
Applying the factorization from \cref{eq:app-cycle-factorization}, and writing \(p_r:=\tr(\rho^r),\) as before, we get
\begin{equation}
\mathbb E_U[X_K(U)^2]
=
\sum_{\tau\in S_{2K}}
\gamma^{(2K)}_\tau(d,m)
\prod_{\omega\in\mathrm{cyc}(\tau\Pi_K)}
p_{|\omega|}.
\label{eq:variance-secondmoment}
\end{equation}

Consequently,
\begin{equation}
\mathrm{Var}_U[X_K]
=
\sum_{\tau\in S_{2K}}
\gamma^{(2K)}_\tau(d,m)
\prod_{\omega\in\mathrm{cyc}(\tau\Pi_K)}
p_{|\omega|}
-
\left[
\sum_{\tau\in S_K}
\gamma^{(K)}_\tau(d,m)
\prod_{\omega\in\mathrm{cyc}(\tau\pi_K)}
p_{|\omega|}
\right]^2 .
\label{eq:variance-general}
\end{equation}

The coefficients \(\gamma\) appearing in Eq.~\eqref{eq:variance-general} are given by the character expansion in \cref{eq:supp-character-coefficients} evaluated at \(K\) and \(2K\), respectively.

\subsection{Example: third projected moment of a noisy GHZ state}

Consider the five-qubit noisy GHZ state
\begin{equation}
\rho
=
(1-p)\,
|\mathrm{GHZ}\rangle\!\langle\mathrm{GHZ}|
+
p\,\frac{\mathbb I}{32},
\qquad
p=0.3,
\label{eq:variance-ghz}
\end{equation}
with Hilbert-space dimension \(d=32\).
We choose a rank-\(m=4\) projector and consider \(K=3\).
The spectrum of \(\rho\) consists of
\(
\lambda_1
=
0.7+\frac{0.3}{32}
=
0.709375,
\)
and
\(
\lambda_2=\cdots=\lambda_{32}
=
\frac{0.3}{32}
=
0.009375.
\)
The required moments are therefore
\begin{equation}
p_r
=
\lambda_1^r
+
31\,\lambda_2^r.
\label{eq:variance-pr}
\end{equation}
Substituting these values into the projected-moment formula yields
\(
\mathbb E_U[X_3]
=
1.4338\times10^{-3}.
\)
Using Eq.~\eqref{eq:variance-secondmoment} evaluated at order \(2K=6\),
one obtains
\(
\mathbb E_U[X_3^2]
=
5.8939\times10^{-6}.
\label{eq:variance-example-second}
\)
Hence
\begin{equation}
\mathrm{Var}_U[X_3]
=
3.8383\times10^{-6},
\label{eq:variance-example-var}
\end{equation}
corresponding to a standard deviation
\(
\sqrt{\mathrm{Var}_U[X_3]}
=
1.96\times10^{-3}.
\)

For \(N\) independent random projections,
the variance of the sample mean scales as
\(
\mathrm{Var}(\overline X_3)
=
\mathrm{Var}_U[X_3]/N.
\)
This variance quantifies the fluctuations arising solely from the choice of Haar-random projection. Additional contributions due to finite measurement statistics and postselection must be added separately when analyzing the total estimator variance.

\subsection{Numerical verification of the variance scaling}

Equation~\eqref{eq:variance-example-var} predicts the variance of the random variable
\[
X_3(U)=\tr(\sigma_U^3)
\]
under Haar-random projections.
This variance characterizes fluctuations arising solely from the random choice of projection subspace and is independent of any measurement shot noise.

To verify the prediction numerically, one may generate independent Haar-random unitaries \(U_1,\cdots,U_N\), evaluate the exact projected moments \(X_3(U_j)\), and form the sample mean
\begin{equation}
\overline X_3
=
\frac1N
\sum_{j=1}^{N}
X_3(U_j),
\label{eq:variance-samplemean-def}
\end{equation}
for which the standard deviation is
\(\mathrm{Std}(\overline X_3)
=
\sqrt{\mathrm{Var}_U[X_3] /N}.\)
For the noisy five-qubit GHZ state of \cref{eq:variance-ghz}, \cref{eq:variance-example-var} yields the prediction
\begin{equation}
\mathrm{Std}(\overline X_3)
=
\frac{1.96\times10^{-3}}{\sqrt N}.
\label{eq:variance-example-prediction}
\end{equation}

\begin{figure}[t]
    \centering
    \includegraphics[width=0.4\linewidth]{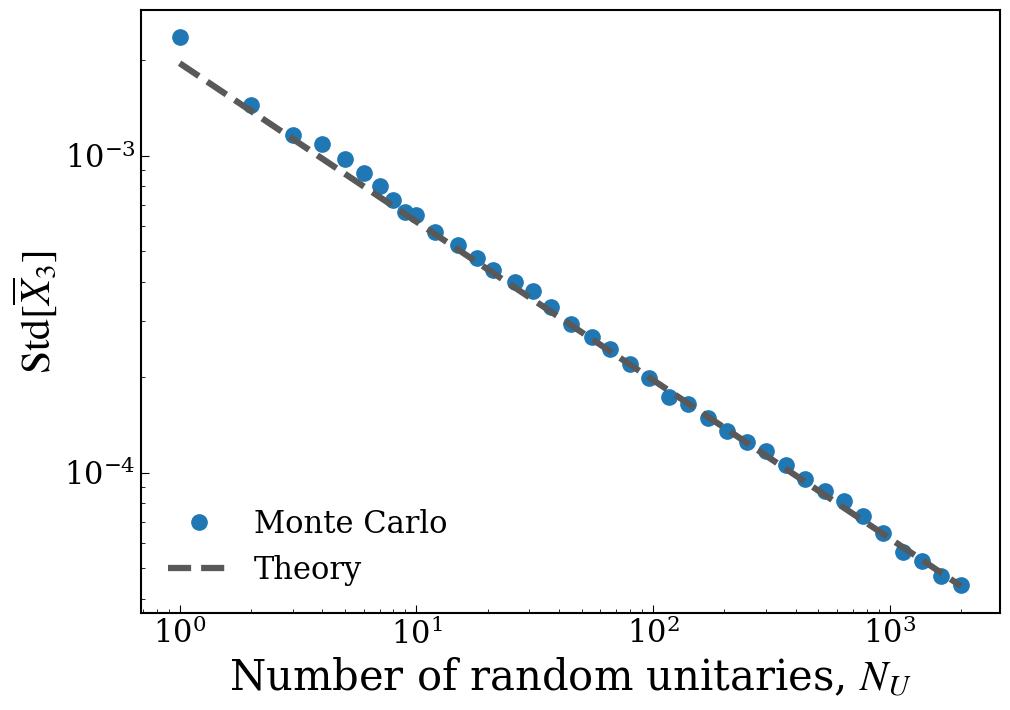}
    \caption{
    Standard deviation of the sample mean
    \(
    \overline X_3
    =
    N_U^{-1}\sum_{j=1}^{N_U}\tr(\sigma_{U_j}^3)
    \)
    for a noisy five-qubit GHZ state (\(p=0.3\)), projection rank \(m=4\), and Hilbert-space dimension \(d=32\).
    Markers show Monte Carlo estimates obtained from independent batches of Haar-random projections, while the dashed curve shows the theoretical prediction
    \(
    \mathrm{Std}(\overline X_3)=\sqrt{\mathrm{Var}_U[X_3]/N_U}.
    \)
    The observed scaling confirms the variance formula derived in this appendix.
    }
    \label{fig:variance_scaling}
\end{figure}

A numerical test may be performed by generating many independent batches of \(N\) Haar-random projections, computing the corresponding sample means \(\overline X_3\), and estimating their empirical standard deviation across batches.
The resulting curve should approach the prediction of Eq.~\eqref{eq:variance-example-prediction} and exhibit the characteristic \(N^{-1/2}\) decay.
This is shown in \cref{fig:variance_scaling}.

It is important to note that this calculation isolates only the fluctuations induced by the random projections themselves.
In the full randomized-projection protocol, additional contributions arise from finite measurement statistics, postselection, and the reconstruction procedure used to recover \(p_3\).
Consequently, the total estimator variance observed in numerical simulations is generally larger than the Haar-projection variance considered here.

\section{Explicit low-order projected moments \label{app:low-order-examples}}

In this section, we derive explicit formulas for low-order projected moments using the coefficient expansion established in Appendix~\ref{sec:supp-proof-theorem1}.
Throughout, we write
\begin{equation}
p_r := \tr(\rho^r),
\qquad
p_1=\tr(\rho)=1,
\end{equation}
and define the projected state
\(
\sigma_U := P_U \rho P_U,
\
P_U = UPU^\dagger,
\)
where \(P\) is a rank-\(m\) projector acting on a \(d\)-dimensional
Hilbert space and \(U\) is Haar-random.
We further use results from previous sections of the Supplemental Material.

\subsection{Two copies}

The partitions of \(2\) are \((2)\) and \((1^2),\) with content polynomials
\(\alpha_{(2)}(x)=x(x+1)\) and \(\alpha_{(1^2)}(x)=x(x-1),\) and with dimensions
\(
f_{(2)}=f_{(1^2)}=1.
\)
The character table of \(S_2\) is
\[
\begin{array}{c|cc}
 & 1^2 & 2 \\
\hline
(2)   & 1 & 1 \\
(1^2) & 1 & -1
\end{array}
\]
where the columns label conjugacy classes.

Using Eq.~\eqref{eq:supp-character-coefficients}, one obtains
\begin{align}
\gamma_{1^2}
&=
\frac12
\left[
\frac{m(m+1)}{d(d+1)}
+
\frac{m(m-1)}{d(d-1)}
\right]
=
\frac{dm^2-m}{d(d^2-1)},
\\[1ex]
\gamma_2
&=
\frac12
\left[
\frac{m(m+1)}{d(d+1)}
-
\frac{m(m-1)}{d(d-1)}
\right]
=
\frac{dm-m^2}{d(d^2-1)}.
\end{align}

Substituting these coefficients into \cref{eq:corollary-sm} yields
\begin{equation}
\mathbb E_U[\tr(\sigma_U^2)]
=
\gamma_{1^2}\,p_2
+
\gamma_2.
\label{eq:loworder-k2}
\end{equation}

Equivalently,
\begin{equation}
p_2
=
\frac{
\mathbb E_U[\tr(\sigma_U^2)]-\gamma_2
}{
\gamma_{1^2}
}.
\label{eq:loworder-k2-inverse}
\end{equation}

\subsection{Three copies}

The partitions of \(3\) are \((3), (2,1)\) and \((1^3).\)
The corresponding content polynomials are \(\alpha_{(3)}(x) = x(x+1)(x+2),\)
\(\alpha_{(2,1)}(x) = x(x^2-1)\) and
\(\alpha_{(1^3)}(x)=x(x-1)(x-2).\)
The irreducible dimensions are
\(\
f_{(3)}=1, f_{(2,1)}=2\) and \(f_{(1^3)}=1.\)
The character table of \(S_3\) is
\[
\begin{array}{c|ccc}
 & 1^3 & 2\,1 & 3 \\
\hline
(3)     & 1 & 1  & 1  \\
(2,1)   & 2 & 0  & -1 \\
(1^3)   & 1 & -1 & 1
\end{array}
\]
where the columns label conjugacy classes.

Using \cref{eq:supp-character-coefficients}, one obtains
\begin{align}
\gamma_{1^3}
&=
\frac16
\left[
\frac{m(m+1)(m+2)}{d(d+1)(d+2)}
+
4\frac{m(m^2-1)}{d(d^2-1)}
+
\frac{m(m-1)(m-2)}{d(d-1)(d-2)}
\right],
\\[1ex]
\gamma_{2,1}
&=
\frac16
\left[
\frac{m(m+1)(m+2)}{d(d+1)(d+2)}
-
\frac{m(m-1)(m-2)}{d(d-1)(d-2)}
\right],
\\[1ex]
\gamma_3
&=
\frac16
\left[
\frac{m(m+1)(m+2)}{d(d+1)(d+2)}
-
2\frac{m(m^2-1)}{d(d^2-1)}
+
\frac{m(m-1)(m-2)}{d(d-1)(d-2)}
\right].
\end{align}

To evaluate Eq.~\eqref{eq:corollary-sm}, let \(\pi=(123)\).
The cycle structure of \(\tau\pi\) determines the corresponding trace monomial through Eq.~\eqref{eq:app-cycle-factorization}.
For \(K=3\), the resulting contributions are proportional to \(1\), \(p_2\), and \(p_3\), yielding

\begin{equation}
\mathbb E_U[\tr(\sigma_U^3)]
=
\gamma_3
+
3\gamma_{2,1}p_2
+
(\gamma_{1^3}+\gamma_3)p_3.
\label{eq:loworder-k3}
\end{equation}

Equation~\eqref{eq:loworder-k3} can be inverted to recover \(p_3\) once an estimate of \(p_2\) is available, illustrating the hierarchical structure of the reconstruction procedure.

\subsection{Four copies}

The partitions of \(4\) are
\(
(4),
(3,1),
(2,2),
(2,1,1)\)
and
\(
(1^4),\)
corresponding to the Young diagrams
\[
\ydiagram{4},
\qquad
\ydiagram{3,1},
\qquad
\ydiagram{2,2},
\qquad
\ydiagram{2,1,1},
\qquad
\ydiagram{1,1,1,1}.
\]

The corresponding content polynomials are
\begin{align}
\alpha_{(4)}(x)
&=
x(x+1)(x+2)(x+3),
\\
\alpha_{(3,1)}(x)
&=
x(x+1)(x+2)(x-1),
\\
\alpha_{(2,2)}(x)
&=
x^2(x+1)(x-1),
\\
\alpha_{(2,1,1)}(x)
&=
x(x+1)(x-1)(x-2),
\\
\alpha_{(1^4)}(x)
&=
x(x-1)(x-2)(x-3).
\end{align}
The irreducible dimensions are obtained from the hook-length formula in \cref{eq:app-hook}, which yields
\(
f_{(4)}=1,
f_{(3,1)}=3,
f_{(2,2)}=2,
f_{(2,1,1)}=3\)
and
\(
f_{(1^4)}=1.
\)
The character table of \(S_4\) is
\[
\begin{array}{c|ccccc}
 & 1^4 & 2\,1^2 & 2^2 & 3\,1 & 4 \\
\hline
(4)       & 1 & 1  & 1  & 1  & 1 \\
(3,1)     & 3 & 1  & -1 & 0  & -1 \\
(2,2)     & 2 & 0  & 2  & -1 & 0 \\
(2,1,1)   & 3 & -1 & -1 & 0  & 1 \\
(1^4)     & 1 & -1 & 1  & 1  & -1
\end{array}
\]

Using \cref{eq:supp-character-coefficients} and \cref{eq:corollary-sm}, one obtains
\begin{equation}
\mathbb E_U[\tr(\sigma_U^4)]
=
A_0
+
A_2 p_2
+
A_{22}p_2^2
+
A_3 p_3
+
A_4 p_4 .
\label{eq:loworder-k4}
\end{equation}
For convenience, we introduce the common denominator
\begin{equation}
D_4
=
d(d-3)(d-2)(d-1)(d+1)(d+2)(d+3).
\label{eq:loworder-k4-denom}
\end{equation}
The coefficients are explicit rational functions of \(d\) and \(m\):
\begin{align}
A_0
&=
\frac{
m(d-m)(d^2-5dm+5m^2+1)
}{
D_4
},
\\[1ex]
A_2
&=
\frac{
2m(d-m)(3d^2m-5dm^2-10d+18m)
}{
D_4
},
\\[1ex]
A_{22}
&=
\frac{
m(d-m)(2d^2m^2+d^2-15dm-3m^2+21)
}{
D_4
},
\\[1ex]
A_3
&=
\frac{
4m(d-m)(d^2m^2+d^2-10dm+m^2+11)
}{
D_4
},
\\[1ex]
A_4
&=
\frac{
m(d^3m^3+5d^3m-20d^2m^2-16d^2+dm^3+65dm-36)
}{
D_4
}.
\end{align}
As a consistency check, in the unprojected limit \(m=d\) all lower-order contributions vanish and one recovers
\begin{equation}
\mathbb E_U[\tr(\sigma_U^4)]
=
p_4.
\qquad (m=d)
\label{eq:loworder-k4-unprojected}
\end{equation}

The appearance of the nonlinear invariant \(p_2^2\) in \cref{eq:loworder-k4} reflects the richer conjugacy-class structure of \(S_4\), which allows products of lower-order spectral moments to arise alongside the genuine fourth-order invariant \(p_4\).

\subsection{Example: \(d=8\), \(m=2\)}

As a concrete illustration, consider three-qubit states (\(d=8\))
projected onto a single qubit (\(m=2\)).
For \(K=2\), \cref{eq:loworder-k2} becomes
\begin{equation}
\mathbb E_U[\tr(\sigma_U^2)]
=
\frac1{42}
+
\frac5{84}p_2.
\label{eq:example-k2}
\end{equation}
For \(K=3\), \cref{eq:loworder-k3} yields
\begin{equation}
\mathbb E_U[\tr(\sigma_U^3)]
=
\frac1{630}
+
\frac1{60}p_2
+
\frac{19}{1260}p_3.
\label{eq:example-k3}
\end{equation}
For \(K=4\), \cref{eq:loworder-k4} becomes
\begin{equation}
\mathbb E_U[\tr(\sigma_U^4)]
=
\frac{1}{27720}
+
\frac{1}{385}p_2
+
\frac{23}{9240}p_2^2
+
\frac{1}{198}p_3
+
\frac{23}{4620}p_4.
\end{equation}

These examples illustrate the hierarchical structure of the projected moments: increasing \(K\) generates progressively richer combinations of spectral invariants, while still permitting reconstruction through the coefficient relations derived above.

\subsection{Algorithmic generation of higher-order projected moments}

For \(K\geqslant 5\), explicit symbolic expressions become lengthier.
Nevertheless, the projected-moment formulas remain straightforward to generate algorithmically from the character representation of the coefficients, as shown in Fig.~\ref{fig:algorithm-projected-moments}.

\begin{figure}[t]
\centering
\fbox{
\begin{minipage}{0.95\columnwidth}

\textbf{Algorithm 1: Generation of projected-moment formulas}

\vspace{0.5em}

\begin{algorithmic}[1]
\Require \(K\in\mathbb N,\; d,m\in\mathbb N, m \leq d\)

\Ensure Explicit formula for \(\mathbb E_U[\tr(\sigma_U^K)]\)

\State Enumerate all partitions \(\lambda\vdash K\).

\State Construct the corresponding Young diagrams.

\State Compute the content polynomials
\[
\alpha_\lambda(x)
=
\prod_{(i,j)\in\lambda}(x+j-i).
\]

\State Compute the dimensions \(f^\lambda\) using the hook-length formula.

\State Evaluate the character table \(\chi_\lambda(\tau)\) of \(S_K\), where \(\tau\in S_K\).

\State Compute the coefficients
\[
\gamma_\tau(d,m)
=
\frac1{K!}
\sum_{\substack{\lambda\vdash K\\
\ell(\lambda)\leqslant d}}
f^\lambda
\frac{\alpha_\lambda(m)}
{\alpha_\lambda(d)}
\chi_\lambda(\tau)
\]
for all relevant conjugacy classes, equivalently for all \(\tau\in S_K\).

\State For each \(\tau\in S_K\), determine the cycle structure of \(\tau\pi\), where \(\pi=(1\,2\,\cdots\,K)\).

\State Associate the trace monomial
\[
\prod_{\omega\in\mathrm{cyc}(\tau\pi)}
p_{|\omega|}
\]
to each permutation \(\tau\).

\State Assemble
\[
\mathbb E_U[\tr(\sigma_U^K)]
=
\sum_{\tau\in S_K}
\gamma_\tau(d,m)
\prod_{\omega\in\mathrm{cyc}(\tau\pi)}
p_{|\omega|}.
\]
\end{algorithmic}

\end{minipage}
}
\caption{
Procedure used to generate explicit projected-moment formulas from the character representation of the coefficients.
}
\label{fig:algorithm-projected-moments}
\end{figure}

The character representation is particularly efficient because the summation runs over partitions of \(K\), whose number grows much more slowly than the \(K!\) elements of the symmetric group.
Consequently, projected moments of moderate order can be generated efficiently on a classical computer and used to construct estimators for higher-order spectral invariants.

\section{Sample Complexity\label{sec:sm-sample-complexity}}

We take a closer look at the sample complexity in terms of the number of required state copies $\rho$.
We first describe how our algorithm is implemented in practice as shown in Fig.~\ref{fig:algorithm-coincidence-random-projection}.
Then we divide this section into two parts.
Firstly, we assume that to estimate $p_K$, the exact values of the state moments of order smaller than $K$ are known. 
Secondly, we consider the stricter scenario in which no information about the lower-order moments is available, and $p_K$ should be estimated by first estimating the lower-order moments hierarchically.

\begin{figure}[b]
\centering
\fbox{
\begin{minipage}{0.95\columnwidth}

\textbf{Algorithm 2: Random projection protocol for estimating \(\{p_k\}_{k=2}^{K}\)}

\vspace{0.5em}

\begin{algorithmic}[1]
\Require Integers \(K\geqslant 2\), \(\frac{(K+2)(K-1)}{2}N_U N_M\) copies of $\rho$ with $n$ qubits, retained subsystem with \(q\) qubits
\Ensure Estimates \(\hat p_2,\cdots,\hat p_K\)

\State Set \(d=2^n\), \(m=2^q\), and \(L=d/m=2^{n-q}\).
\State Precompute the reconstruction relations
\[
\bar\sigma_k^{(d,m)}
=
\gamma^{(k)}(d,m)p_k
+
F_k(p_1, p_2,\cdots,p_{k-1}),
\qquad k=2,\cdots,K,
\]
with \(p_1=1\).

\For{\(k=2,\cdots,K\)}
    \For{\(s=1,\cdots,N_U\)}
        \State Randomly choose a unitary \(U_s\) from the (approximated) Haar-random unitary ensemble \(\mathcal E\).

        \For{\(j=1,\cdots,N_M\)}
            \State Prepare \(k\) copies of \(\rho\).
            \State Apply \(U_s\) to each copy.
            \State Measure \(n-q\) qubits of each copy in the computational basis.
            \State Record the projection outcomes \(y_1,\cdots,y_k\in\{1,\cdots,L\}\).

            \If{\(y_1=\cdots=y_k\)}
                \State Keep the remaining \(q\) qubits of all \(k\) copies.
                \State Perform the generalized swap test on the retained \(k\) compressed copies.
                \State Record the outcome \(Y_{s,j}^{(k)}\in\{+1,-1\}\).
            \Else
                \State Set \(Y_{s,j}^{(k)}=0\).
            \EndIf
        \EndFor
    \EndFor

    \State Estimate the Haar-averaged projected moment by
    \[
    \hat{\bar\sigma}_k^{(d,m)}
    =
    \frac{1}{L N_U N_M}
    \sum_{s=1}^{N_U}
    \sum_{j=1}^{N_M}
    Y_{s,j}^{(k)} .
    \]
\EndFor

\For{\(k=2,\cdots,K\)}
    \State Compute recursively
    \[
    \hat p_k
    =
    \frac{
    \hat{\bar\sigma}_k^{(d,m)}
    -
    F_k(\hat p_1,\cdots,\hat p_{k-1})
    }
    {
    \gamma^{(k)}(d,m)
    },
    \qquad \hat p_1=1.
    \]
\EndFor

\State \Return \(\hat p_2,\cdots,\hat p_K\).

\end{algorithmic}

\end{minipage}
}
\caption{
Random projection protocol for estimating $p_2,\cdots,p_K$.
}
\label{fig:algorithm-coincidence-random-projection}
\end{figure}

\subsection{Known lower-order moments}

In this scenario, we have the exact value of $F_K(p_1,\cdots,p_{K-1})$.
From Corollary~1:
\begin{equation}
    \bar\sigma_K(\rho) = 
\sum_{\tau\in S_K}
\gamma_\tau(d,m)
\prod_{j=1}^{\mathrm{c}(\tau\pi)}
\tr\!\left ( \rho^{\nu_j} \right )=
    \gamma^{(K)}\, p_K
    +
    F_K\!\left(p_1,\cdots,p_{K-1}\right),
\end{equation}
where $\bar\sigma_K(\rho) = \bar \sigma^{(d,m)}_K$ and $\gamma^{(K)}=\gamma^{(K)}(d,m)$, with $(d,m)$ omitted.
Therefore:
\begin{equation}
|\hat p_K-p_K|
=
\frac{
|\hat{\bar{\sigma}}_K-\bar{\sigma}_K|
}{
|\gamma^{(K)}|
},
\end{equation}
where the hat notation denotes an estimator of the corresponding quantity.
Suppose we compress the system to the reduced Hilbert space $(\mathbb{C}^{2^q})^{\otimes K}$, i.e., we measure $n-q$ qubits before applying the generalized swap test. 
Then there will be $L=2^{n-q}$ branches of measurement outcomes. 
For each $y\in\{1,2,\cdots,L\}$ and fixed $U$, we have the corresponding projector $P_y$ and we denote $\sigma_{U,y}=U P_y U^{\dagger} \rho U P_y U^{\dagger}$.
Note that $\sigma_{U,y}$ is not normalized.
The normalization factor is $\mathrm{Pr}_{U,y} = \tr(\sigma_{U,y})$.
Then the corresponding density matrix is $\rho_{U,y}=\mathrm{Pr}^{-1}_{U,y}\sigma_{U,y}$.
Note that for the generalized swap test, $\mathbb{E}_{\mathrm{shots}}[S|U,y]=\tr(\rho^K_{U,y})$, where $S\in\{+1,-1\}$ denotes the outcome corresponding to measuring the ancillary qubit in $\ket{0}$ and $\ket{1}$, respectively.
Then $\tr(\sigma^K_{U,y}) = \mathrm{Pr}^K_{U,y}\tr(\rho^K_{U,y}) = \mathrm{Pr}^K_{U,y}\mathbb{E}_{\mathrm{shots}}[S|U,y] $.
Note that $\mathrm{Pr}^K_{U,y}$ is exactly the probability that all \(K\) copies are projected onto the subspace associated with $P_y$.
We denote $Y\in\{+1,0,-1\}$. 
The outcomes $\{+1,-1\}$ are defined as for $S$, while $\{0\}$ corresponds to a failed run, i.e., the \(K\) input states are not projected onto the same subspace.
Since we are interested in successful coincidence events across all branches, rather than conditioning on a particular branch, we have
\begin{equation}
    \mathbb{E}_{\mathrm{shots}}[Y|U]=\sum_{y=1}^{L}\tr(\sigma^K_{U,y}).
\end{equation}
And naturally:
\begin{equation}
    \mathbb{E}_{\mathrm{shots}, U}[Y]=\sum_{y=1}^{L}\mathbb{E}_{U}[\tr(\sigma^K_{U,y})]=L\mathbb{E}_{U}[\tr(\sigma^K_U)] = L\bar{\sigma}_K,
\end{equation}
i.e.,
\begin{equation}
    \bar{\sigma}_K = \frac{1}{L}\mathbb{E}_{\mathrm{shots}, U}[Y].
\end{equation}
Therefore, given $N_U$ sampled $U$ and $N_M$ executions per $U$, we denote the value of $Y$ at $a$-th $U$ and $b$-th execution as $Y_{a,b}$. We then have:
\begin{equation}
\begin{split}
    |\hat{\bar{\sigma}}_K-\bar{\sigma}_K| &= \left| \frac{1}{L N_U N_M}\sum_{a=1}^{N_U}\sum_{b=1}^{N_M}Y_{a,b} - \bar{\sigma}_K \right| \\
    &=\left| \frac{1}{L N_U N_M}\sum_{a=1}^{N_U}\sum_{b=1}^{N_M}(\mu(U_a)+Y_{a,b}-\mu(U_a)) - \bar{\sigma}_K \right| \\
    &=\left| \left(\frac{1}{L N_U}\sum_{a=1}^{N_U}\mu(U_a) - \bar{\sigma}_K\right) + \frac{1}{L N_U N_M}\sum_{a=1}^{N_U}\sum_{b=1}^{N_M}(Y_{a,b}-\mu(U_a)) \right| \\
    &\leqslant \left| \frac{1}{L N_U}\sum_{a=1}^{N_U}\mu(U_a) - \bar{\sigma}_K \right| + \left| \frac{1}{L N_U N_M}\sum_{a=1}^{N_U}\sum_{b=1}^{N_M}(Y_{a,b}-\mu(U_a)) \right|,
\end{split}
\end{equation}
where $\mu(U_a)=\mathbb{E}_{\mathrm{shots}}[Y_{a,b}|U_a]=\sum_{y=1}^{L}\tr(\sigma_{U_a,y}^K)$ and naturally $\mathbb{E}_U[\mu(U)]=L\bar{\sigma}_K$.
The expression written above is to explicitly show that the estimation error of $\bar{\sigma}_K$ has two kinds of errors: 1) estimated with limited sampled $U$, even when $N_M\rightarrow \infty$, and 2) estimated with limited shots, which is limited by both $N_M$ and $N_U$.
Therefore, increasing $N_U$ is particularly important, since it reduces both the unitary-sampling error and the shot-noise contribution.

From the law of total variance, we then have:
\begin{equation}
    \mathrm{Var}(\hat{\bar{\sigma}}_K) = \mathrm{Var}_\mathcal{U}(\mathbb{E}_{\mathrm{shots}}[\hat{\bar{\sigma}}_K|\mathcal{U}]) + \mathbb{E}_\mathcal{U}[\mathrm{Var}_{\mathrm{shots}}(\hat{\bar{\sigma}}_K|\mathcal{U})],
\end{equation}
where $\mathcal{U}$ denotes the ensemble of $U_1,\cdots U_{N_U}$. Then:
\begin{equation}
    \mathbb{E}_{\mathrm{shots}}[\hat{\bar{\sigma}}_K|\mathcal{U}]=\mathbb{E}_{\mathrm{shots}}\left[ \frac{1}{L N_U N_M}\sum_{a=1}^{N_U}\sum_{b=1}^{N_M}Y_{a,b} \mid \mathcal{U} \right] = \frac{1}{LN_U N_M}\sum_{a=1}^{N_U}\sum_{b=1}^{N_M}\mathbb{E}_{\mathrm{shots}}[Y_{a,b}|U_a]=\frac{1}{LN_U}\sum_{a=1}^{N_U}\mu(U_a).
\end{equation}
Therefore:
\begin{equation}
    \mathrm{Var}_\mathcal{U}(\mathbb{E}_{\mathrm{shots}}[\hat{\bar{\sigma}}_K|\mathcal{U}]) = \frac{1}{L^2 N_U^2} N_U \mathrm{Var}_{U}(\mu(U)) = \frac{1}{L^2 N_U} \mathrm{Var}_{U}(\mu(U)).
\end{equation}
For the second term, since:
\begin{equation}
    \mathrm{Var}_{\mathrm{shots}}(\hat{\bar{\sigma}}_K|\mathcal{U}) = \mathrm{Var}_{\mathrm{shots}}\left( \frac{1}{L N_U N_M}\sum_{a=1}^{N_U}\sum_{b=1}^{N_M}Y_{a,b}  \mid \mathcal{U} \right) = \frac{1}{L^2 N_U^2 N_M}\sum_{a=1}^{N_U}\mathrm{Var}_{\mathrm{shots}}(Y | U_a),
\end{equation}
then:
\begin{equation}
    \mathbb{E}_\mathcal{U}[\mathrm{Var}_{\mathrm{shots}}(\hat{\bar{\sigma}}_K|\mathcal{U})] = \frac{1}{L^2N_U N_M}\mathbb{E}_{U}[\mathrm{Var}_{\mathrm{shots}}(Y | U)].
\end{equation}
Since $Y^2\in\{0,1\}$ denotes the failure or success of the simultaneous identical projections,
therefore $\mathbb{E}_{\mathrm{shots}}(Y^2 | U) = \sum_{y=1}^{L}\mathrm{Pr}_{U,y}^K$. 
Then:
\begin{equation}
    \mathrm{Var}_{\mathrm{shots}}(Y | U) = \mathbb{E}_{\mathrm{shots}}(Y^2 | U) - \mathbb{E}^2_{\mathrm{shots}}(Y | U) = \sum_{y=1}^{L}\mathrm{Pr}_{U,y}^K - \mu^2(U)\leqslant\sum_{y=1}^{L}\mathrm{Pr}_{U,y}^K.
\end{equation}
As $\mu(U)=\sum_{y=1}^{L}\tr(\sigma_{U,y}^K)=\sum_{y=1}^{L}\mathrm{Pr}_{U,y}^K\tr(\rho^K_{U,y})\leqslant\sum_{y=1}^{L}\mathrm{Pr}_{U,y}^K$, 
we would like to find an approximate scaling of $\mathrm{Pr}_{U,y}^K$.
Note that $\mathrm{Pr}_{U,y}=\tr(UP_yU^{\dagger}\rho)=\tr(P_{U,y}\rho)$.

We start from the pure state $\rho=\ket{\psi}\bra{\psi}$.
For simplicity, we suppose for one certain branch $y$, $P_{U,y}$ is a rank-$m$ projector such that $P_y=\sum_{j=0}^{m-1}\ket{j}\bra{j}$.
Since $\tr(P_{U,y}\rho)=\bra{\psi}UP_yU^{\dagger}\ket{\psi}=\bra{\phi}P_y\ket{\phi}$ where $\ket{\phi}$ is a Haar-random pure state, we can write $\ket{\phi}$ in the unnormalized computational basis with each corresponding factor sampled from $g=\frac{X+iY}{\sqrt{2}}$ and $X,Y\sim \mathcal{N}(0,1)$.
Then we define $R=|g|^2=\frac{X^2+Y^2}{2}$, and start calculating the cumulative density function (CDF) of $R$, i.e., $\mathrm{Pr}(R\leqslant r)$ for $r\geqslant0$.
Clearly, $\mathrm{Pr}(R\leqslant r)=\mathrm{Pr}\left(\frac{X^2+Y^2}{2}\leqslant r \right) = \mathrm{Pr}\left(X^2+Y^2\leqslant 2r \right)$.
This means $(X,Y)$ lies inside a circle of radius $\sqrt{2r}$.
The joint Gaussian density of $(X,Y)$ is $f_{X,Y}(x,y)=\frac{1}{2\pi} e^{-\frac{x^2+y^2}{2}}$, then:
\begin{equation}
    \mathrm{Pr}(X^2+Y^2\leqslant 2r) = \frac{1}{2\pi}\iint_{x^2+y^2\leqslant 2r} e^{-\frac{x^2+y^2}{2}}dxdy.
\end{equation}
Switching to polar coordinates with $x=\rho\cos\theta$, $y=\rho\sin\theta$ and $dxdy=\rho d\rho d\theta$, we find that
\begin{equation}
    \mathrm{Pr}(X^2+Y^2\leqslant 2r) = \frac{1}{2\pi} \int_{0}^{2\pi}\int_{0}^{\sqrt{2r}} \rho e^{-\frac{\rho^2}{2}}d\rho d\theta = \int_{0}^{\sqrt{2r}}\rho e^{-\frac{\rho^2}{2}}d\rho = 1-e^{-r},
\end{equation}
which is exactly the CDF of the exponential distribution $\sim\mathrm{Exp}(1)$.
Since:
\begin{equation}
    \mathrm{Pr}_{U,y}=\bra{\phi}P\ket{\phi}=\frac{\sum_{i=0}^{m-1}|g_i|^2}{\sum_{i=0}^{m-1}|g_i|^2 + \sum_{i=m}^{d-1}|g_i|^2}=\frac{A}{A+B},
\end{equation}
and $|g_i|^2\sim\mathrm{Exp}(1)$, then $A\sim \mathrm{Gamma}(m,1)$ and $B\sim \mathrm{Gamma}(d-m,1)$.
Since $A$ and $B$ are independently distributed, then:
\begin{equation}
    \mathrm{Pr}_{U,y} = \frac{A}{A+B} \sim \mathrm{Beta}(m, d-m).
\end{equation}
Therefore:
\begin{equation}
    \mathbb{E}_{U}[\mathrm{Pr}_{U,y}^K] = \prod_{j=0}^{K-1}\frac{m+j}{d+j} = \Upsilon(m, d, K).
\end{equation}
For the variance, we have
\begin{equation}
    \mathrm{Var}_{U}(\mathrm{Pr}_{U,y}^K) = \Upsilon(m, d, 2K) - \Upsilon^2(m, d, K).
\end{equation}
Since for pure state $\rho=\ket{\psi}\bra{\psi}$, 
\begin{equation}
    \mu(U) = \sum_{y=1}^{L}\mathrm{Pr}_{U,y}^K \tr(\rho^K_{U,y}) = \sum_{y=1}^{L}\mathrm{Pr}_{U,y}^K,
\end{equation}
therefore $\mathbb{E}_U[\mu(U)]=\mathbb{E}_U\left[\sum_{y=1}^{L}\mathrm{Pr}_{U,y}^K\right]=L\Upsilon(m,d,K)$.
Now compute $\mathbb{E}_U[\mu^2(U)]$, since:
\begin{equation}
    \mu^2(U) = \sum_{y=1}^{L}\mathrm{Pr}_{U,y}^{2K} + \sum_{y\neq y'}\mathrm{Pr}_{U,y}^{K}\mathrm{Pr}_{U,y'}^{K},
\end{equation}
then:
\begin{equation}
    \mathbb{E}_U[\mu^2(U)] = \sum_{y=1}^{L}\mathbb{E}_U[\mathrm{Pr}_{U,y}^{2K}] + \sum_{y\neq y'}\mathbb{E}_U[\mathrm{Pr}_{U,y}^{K}\mathrm{Pr}_{U,y'}^{K}].
\end{equation}
Since $\sum_{y=1}^L\mathrm{Pr}_{U,y}=1$ and $\mathrm{Pr}_{U,y} \sim \mathrm{Beta}(m, d-m)$,  we assume the joint distribution $(\mathrm{Pr}_{U,1}, \cdots,\mathrm{Pr}_{U,L})\sim \mathrm{Dirichlet}(m,\cdots,m)$. 
Therefore, from the moments of Dirichlet distribution, we have:
\begin{equation}
    \mathbb{E}_U[\mathrm{Pr}_{U,y}^{K}\mathrm{Pr}_{U,y'}^{K}] = \frac{\Gamma(d)\Gamma^2(m+K)}{\Gamma(d+2K)\Gamma^2(m)}=\frac{\prod_{j=0}^{K-1}(m+j)^2}{\prod_{j=0}^{2K-1}(d+j)}.
\end{equation}
Therefore:
\begin{equation}
    \mathbb{E}_U[\mu^2(U)] = L\Upsilon(m,d,2K) + L(L-1)\frac{\prod_{j=0}^{K-1}(m+j)^2}{\prod_{j=0}^{2K-1}(d+j)}.
\end{equation}
Then:
\begin{equation}
    \mathrm{Var}(\mu(U)) = \mathbb{E}_U[\mu^2(U)]  - \mathbb{E}^2_U[\mu(U)] = L\Upsilon(m,d,2K)+ L(L-1)\frac{\prod_{j=0}^{K-1}(m+j)^2}{\prod_{j=0}^{2K-1}(d+j)} - L^2 \Upsilon^2(m,d,K)
\end{equation}

For mixed state $\rho$, we eigendecompose $\rho$ as $\rho=\sum_{i}\lambda_i\ket{\psi_i}\bra{\psi_i}$ with $\lambda_i\geqslant0$ and $\sum_{i}\lambda_i=1$.
Then $\mathrm{Pr}_{U,y}=\tr(P_{U,y}\rho) = \sum_{i}\lambda_i\bra{\phi_i} P_y \ket{\phi_i}$.
Since $x \rightarrow x^K$ is convex on $x\geqslant 0$, we then obtain from Jensen's inequality:
\begin{equation}
    \mathrm{Pr}_{U,y}^K = \left(\sum_{i}\lambda_i\bra{\phi_i} P_y \ket{\phi_i}\right)^K \leqslant\sum_{i}\lambda_i \left( \bra{\phi_i} P_y \ket{\phi_i} \right)^K,
\end{equation}
and therefore
\begin{equation}
    \mathbb{E}_{U}[\mathrm{Pr}_{U,y}^K] \leqslant \sum_{i}\lambda_i \mathbb{E}_{U}[\left( \bra{\phi_i} P \ket{\phi_i} \right)^K] = \Upsilon(m,d,K).
\end{equation}
For the variance of $\mu(U)$, we have:
\begin{equation}
\begin{split}
    & \mathrm{Var}_U(\mu(U))\leqslant \mathbb{E}_U[\mu^2(U)] \leqslant \mathbb{E}_U\left[\left( \sum_{y=1}^{L}\mathrm{Pr}^K_{U,y} \right)^2\right] \leqslant \mathbb{E}_U\left[ \left(\sum_{i}\lambda_i\sum_{y=1}^{L}  (\bra{\phi_i}P_y\ket{\phi_i})^K\right)^2 \right] \\
\leqslant & \sum_{i}\lambda_i\mathbb{E}_U\left[ \left(\sum_{y=1}^{L}  (\bra{\phi_i}P_y\ket{\phi_i})^K\right)^2 \right] = L\Upsilon(m,d,2K) + L(L-1)\frac{\prod_{j=0}^{K-1}(m+j)^2}{\prod_{j=0}^{2K-1}(d+j)}.
\end{split}
\end{equation}
Therefore:
\begin{equation}
\begin{split}
    &\mathrm{Var}(\hat{\bar{\sigma}}_K)\leqslant\frac{1}{L^2 N_U}\mathrm{Var}_U(\mu(U)) + \frac{1}{L^2N_UN_M}\mathbb{E}_{U}\left[\sum_{y=1}^{L}\mathrm{Pr}^K_{U,y}\right] \\
\leqslant&\frac{\Upsilon(m,d,2K) + (L-1)\frac{\prod_{j=0}^{K-1}(m+j)^2}{\prod_{j=0}^{2K-1}(d+j)}}{LN_U} + \frac{1}{L N_U N_M}\Upsilon(m,d,K).
\end{split}
\end{equation}
For simplicity, we approximate $\Upsilon(m,d,2K)\sim L^{-2K}, \Upsilon(m,d,K)\sim L^{-K}, \frac{\prod_{j=0}^{K-1}(m+j)^2}{\prod_{j=0}^{2K-1}(d+j)}\sim L^{-2K}$, 
since $K$ is considered to be relatively small compared with $m$ and $d$.
Therefore:
\begin{equation}
    \mathrm{Var}(\hat{\bar{\sigma}}_K) \lesssim \frac{L^{-2K}}{N_U} + \frac{L^{-K-1}}{N_U N_M}.
\end{equation}
Therefore, with Chebyshev's inequality:
\begin{equation}
    \Pr(|\hat{\bar{\sigma}}_K-\bar{\sigma}_K|\geqslant\epsilon_{\sigma})\leqslant\frac{\frac{L^{-2K}}{N_U} + \frac{L^{-K-1}}{N_U N_M}}{\epsilon_{\sigma}^2}.
\end{equation}
Then:
\begin{equation}
    |\hat{\bar{\sigma}}_K-\bar{\sigma}_K|\sim O\left(\sqrt{\frac{L^{-2K}}{N_U} + \frac{L^{-K-1}}{N_U N_M}}\right),
\end{equation}
and:
\begin{equation}
    |\hat{p}_K-p_K|\sim O\left(\frac{1}{|\gamma^{(K)}|}\sqrt{\frac{L^{-2K}}{N_U} + \frac{L^{-K-1}}{N_U N_M}}\right).
\end{equation}
We approximate $|\gamma^{(K)}|\sim L^{-K}$, and correspondingly find for the statistical error that
\begin{equation}
    |\hat{p}_K - p_K| = \epsilon_{p} = \epsilon_{\mathrm{stat}} \sim O\left( \sqrt{\frac{1}{N_U}+ \frac{L^{K-1}}{N_U N_M}} \right).
\end{equation}
Then it is sufficient to choose:
\begin{equation}
    N_U\sim O\left( \frac{1}{\epsilon_{\mathrm{stat}}^2}\left( 1+\frac{L^{K-1}}{N_M} \right) \right),
\end{equation}
and the required total number of copies $N=KN_UN_M$ is approximately:
\begin{equation}
    N \sim O\left( \frac{K}{\epsilon_{\mathrm{stat}}^2}(N_M + L^{K-1}) \right).
\end{equation}
This estimate only captures the statistical error from the finite numbers of random circuits and measurement repetitions. 
It assumes that the implemented local brickwork circuit ensemble is sufficiently close to the target Haar ensemble, 
so that the corresponding approximation bias is negligible. 
More generally, if the brickwork ensemble induces a bias:
\begin{equation}
    \epsilon_{\mathrm{bw}} = |\mathbb{E}_{\mathrm{brickwork}}[\hat{p}_K]-p_K|,
\end{equation}
Then the total error should be decomposed as:
\begin{equation}
    |\hat{p}_K - p_K| \leqslant |\hat{p}_K - \mathbb{E}_{\mathrm{brickwork}}[\hat{p}_K]| + |\mathbb{E}_{\mathrm{brickwork}}[\hat{p}_K] - p_K|.
\end{equation}
Hence,
\begin{equation}
    \epsilon_p \leqslant \epsilon_{\mathrm{stat}} + \epsilon_{\mathrm{bw}}.
\end{equation}
If one requires a final accuracy $\epsilon_p$, then the statistical budget should be chosen as:
\begin{equation}
    \epsilon_{\mathrm{stat}} = \epsilon_p - \epsilon_{\mathrm{bw}}.
\end{equation}
Assuming $\epsilon_{\mathrm{bw}} \ll \epsilon_p$, then:
\begin{equation}
    N \sim O\left( \frac{K}{\epsilon_{\mathrm{stat}}^2}(N_M + L^{K-1}) \right) =  O\left( \frac{K}{(\epsilon_p - \epsilon_{\mathrm{bw}})^2}(N_M + L^{K-1}) \right).
\end{equation}

\subsection{Unknown lower-order moments}
\begin{figure}
    \centering
    \includegraphics[width=1.0\linewidth]{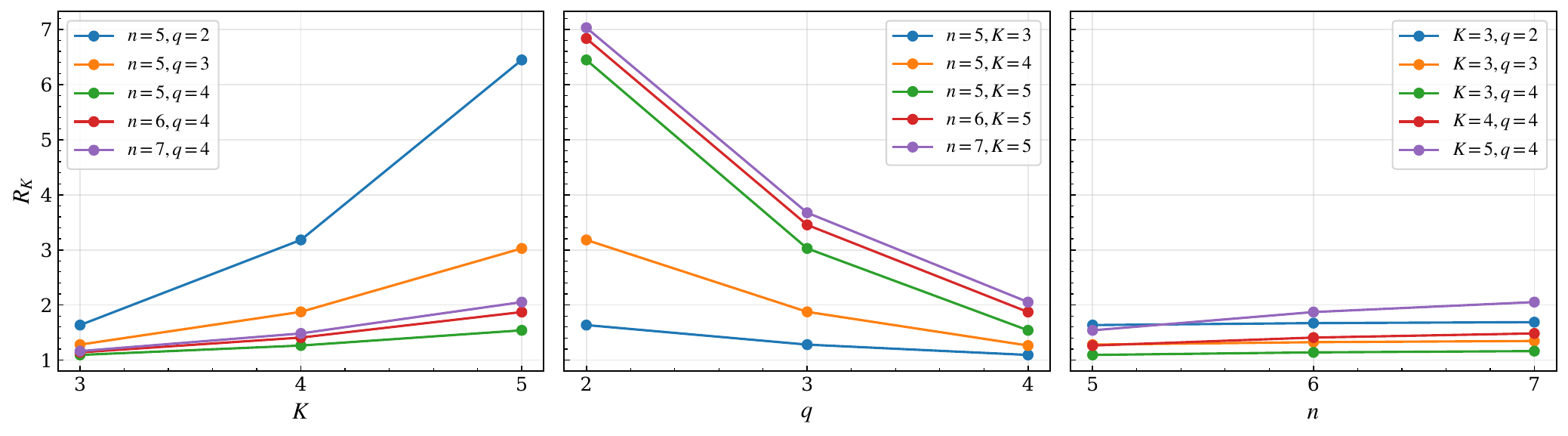}
    \caption{The error amplification factor $R_K$ as given by \cref{eq:RK-amplification} under different $K$, $q$ and $n$.}
    \label{fig:mean_value_bound}
\end{figure}

In this scenario, $F_K(p_1,\cdots,p_{K-1})$ is actually unknown. 
$p_K$ should then be estimated by first estimating the lower-order moments hierarchically.
Therefore:
\begin{equation}
    |\hat{p}_K-p_K|\leqslant\frac{|\hat{\bar{\sigma}}_K-\bar{\sigma}_K|}{|\gamma^{(K)}|} + \frac{|F_K - \hat{F}_K|}{|\gamma^{(K)}|},
\end{equation}
where we denote $F_K=F_K(p_1,\cdots,p_{K-1})$ and $\hat{F}_K=F_K(\hat{p}_1,\cdots,\hat{p}_{K-1})$.
As derived before, the statistical error of the first term is of order $ O\left( \sqrt{\frac{1}{N_U}+ \frac{L^{K-1}}{N_U N_M}} \right)$.
The second term can be bounded using the mean value theorem such that:
\begin{equation}
    |F_K - \hat{F}_K| \leqslant \sum_{k=2}^{K-1}|\hat{p}_k - p_k| \sup_{\mathbf{x}}\left| \frac{\partial F_K}{\partial p_k}(\mathbf{x}) \right|,
\end{equation}
where $\mathbf{x}$ lies on the line segment between $(p_2,\cdots,p_{K-1})$ and $(\hat{p}_2,\cdots,\hat{p}_{K-1})$. 
Here we assume that each reconstructed moment estimate is clipped to the physical interval $[0,1]$.
We omit $p_1$ since it is fixed as $1$.
Given that
\begin{equation}
\bar\sigma_K(\rho) = 
\sum_{\tau\in S_K}
\gamma_\tau(d,m)
\prod_{j=1}^{\mathrm{c}(\tau\pi)}
\tr\!\left ( \rho^{\nu_j} \right ),
\end{equation}
we know that
\begin{equation}
    \gamma^{(K)} = \sum_{\substack{\tau \in S_{K} \\ \text{$\tau\pi$ is a $K$-cycle}}}\gamma_{\tau}(d,m),
\end{equation}
and
\begin{equation}
    F_K = \sum_{\substack{\tau \in S_{K} \\ \text{$\tau\pi$ is NOT a $K$-cycle}}}\gamma_{\tau}(d,m) \prod_{j=1}^{c(\tau \pi)} p_{\nu_j}.
\end{equation}
Let $a_s(\tau)$ denote the number of cycles of $\tau\pi$ with length $s$.
Take the partial derivative of $F_K$:
\begin{equation}
\begin{split}
    \frac{\partial F_K}{\partial p_k} =& \sum_{\substack{\tau \in S_{K} \\ \text{$\tau\pi$ is NOT a $K$-cycle}}} \gamma_{\tau}(d,m)  \frac{\partial}{\partial p_k} \left( \prod_{s=2}^{K-1}p_s^{a_s(\tau)} \right) \\ 
    =& \sum_{\substack{\tau \in S_{K} \\ \text{$\tau\pi$ is NOT a $K$-cycle}}}\gamma_{\tau}(d,m)a_k(\tau) p_k^{a_k(\tau)-1}\prod_{s=2,s\neq k}^{K-1}p_s^{a_s(\tau)}.    
\end{split}
\end{equation}
Compute the upper bound by considering $p_k\leqslant1$ for all $k$, then:
\begin{equation}
    \frac{|F_K - \hat{F}_K|}{|\gamma^{(K)}|}\leqslant \sum_{k=2}^{K-1}|\hat{p}_k - p_k|\frac{\sum_{\substack{\tau \in S_{K} \\ \text{$\tau\pi$ is NOT a $K$-cycle}}}\left|\gamma_{\tau}(d,m)\right|a_k(\tau) }{\left|\sum_{\substack{\tau \in S_{K} \\ \text{$\tau\pi$ is a $K$-cycle}}}\gamma_{\tau}(d,m)\right|} = \sum_{k=2}^{K-1}|\hat{p}_k - p_k| A_{K,k} \leqslant \max_{2\leqslant k \leqslant K-1}|\hat{p}_k-p_k| \cdot\sum_{k=2}^{K-1} A_{K,k} .
\end{equation}
Since $L\geqslant 1$ and $k\leqslant K$, the direct statistical error at order $k$ is bounded by the $K$-order scale,
i.e., $\sqrt{\frac{1}{N_U}+\frac{L^{k-1}}{N_U N_M}}\leqslant\sqrt{\frac{1}{N_U}+\frac{L^{K-1}}{N_U N_M}}$,
and therefore:
\begin{equation}
    |\hat{p}_K - p_K| \sim O\left( R_K\sqrt{\frac{1}{N_U} + \frac{L^{K-1}}{N_UN_M}} \right),
\end{equation}
where,
\begin{equation}\label{eq:RK-amplification}
    R_2 = 1, \ \ \ \ R_K = 1 + \sum_{k=2}^{K-1}A_{K,k} R_k
\end{equation}
This upper bound tracks the error amplification $R_K$ through the hierarchical moment estimation.
The scaling of this factor may not be analytically tractable in general. 
However, it can be evaluated directly from the polynomial coefficients, and its behavior is shown numerically in Fig.~\ref{fig:mean_value_bound}. 
The amplification factor increases with $K$ and decreases with the retained subsystem size $q$, while remaining moderate in the parameter regimes considered.
Thus, for small $K$ and not too aggressive compression, the recursive reconstruction leads to only controlled, approximately linear propagation of lower-moment errors.
Therefore, in the small-$K$ regime, $R_K$ remains moderate and can be absorbed into the $\mathrm{poly}(K)$ prefactor.
For simplicity, we assume that we also use $kN_UN_M$ copies to estimate $p_k$ for every $k=2,\cdots,K-1$.
Therefore:
\begin{equation}
    |\hat{p}_K-p_K| \sim  O\left( \mathrm{poly}(K)\sqrt{\frac{1}{N_U}+ \frac{L^{K-1}}{N_U N_M}} \right).
\end{equation}
Following a similar approach, we also find that
\begin{equation}
    N_U \sim O \left( \frac{\mathrm{poly}(K)}{\epsilon_{\mathrm{stat}}^2}\left( 1+\frac{L^{K-1}}{N_M} \right) \right),
\end{equation}
and since in this case $N = \frac{(K+2)(K-1)}{2}N_U N_M$, we have:
\begin{equation}
    N \sim O\left( \frac{\mathrm{poly}(K)}{\epsilon_{\mathrm{stat}}^2}(N_M + L^{K-1})  \right).
\end{equation}
The brickwork circuit approximation error can also be incorporated here in a similar way.

Overall, if one needs to estimate $p_K$ with absolute error $\epsilon_p$, it requires:
\begin{equation}
    N \sim O\left( \mathrm{poly}(K) \frac{N_M + L^{K-1}}{(\epsilon_p - \epsilon_{\mathrm{bw}})^2} \right),
\end{equation}
with $N=KN_UN_M$ when lower-order moments are known and $N=\frac{(K+2)(K-1)}{2}N_U N_M$ when lower-order moments are unknown.

\section{Comparison with local randomized measurements \label{sec:q0-comparison}}

\begin{figure*}[t]
  \centering

  \subfloat[]{%
    \includegraphics[width=0.5\linewidth]{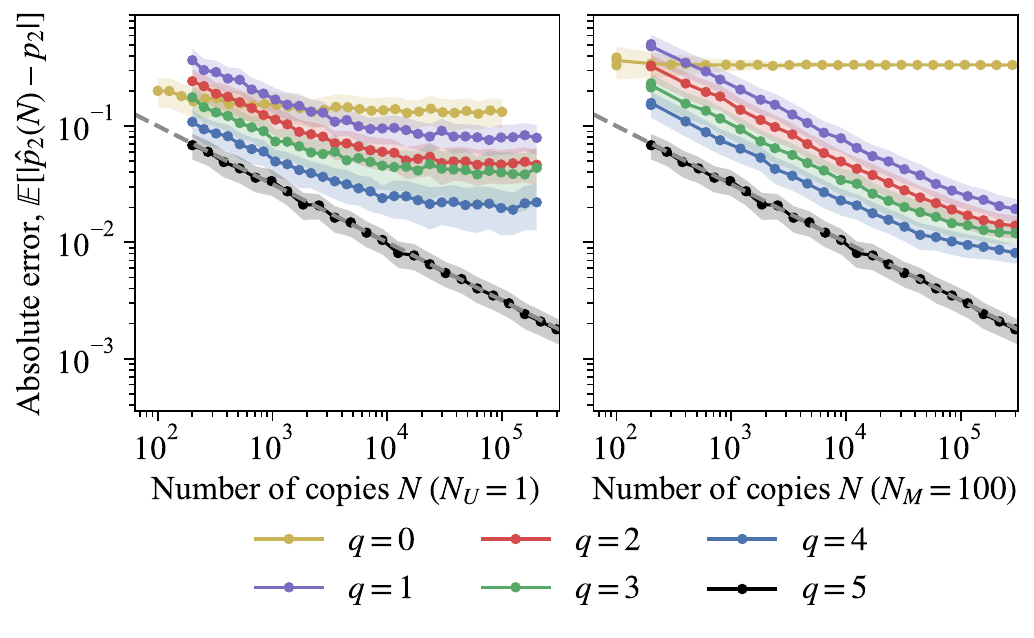}%
    \label{fig:N_vs_error_K2}%
  }
  \hfill
  \subfloat[]{%
    \includegraphics[width=0.5\linewidth]{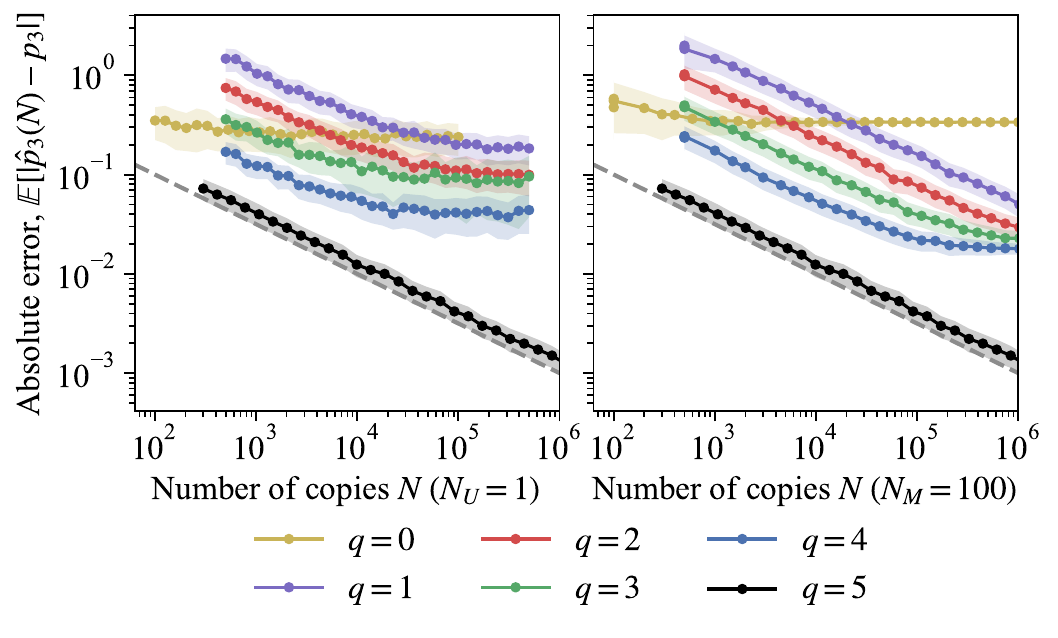}%
    \label{fig:N_vs_error_K3}%
  }

  \vspace{0.3cm}

  \makebox[\linewidth][c]{%
    \subfloat[]{%
      \includegraphics[width=0.5\linewidth]{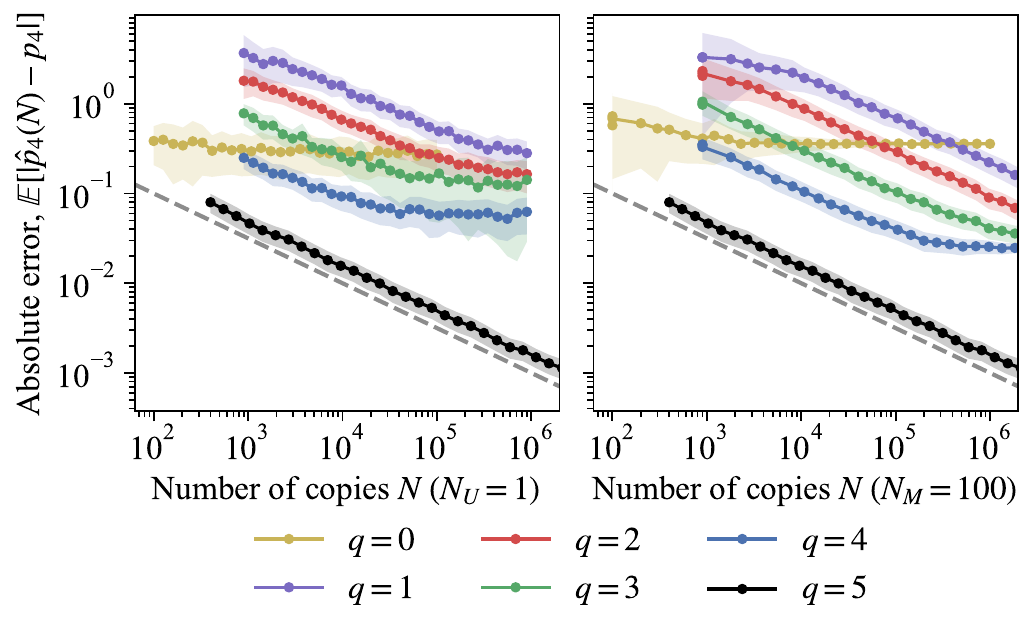}%
      \label{fig:N_vs_error_K4}%
    }%
  }

  \caption{
  Mean absolute estimation error $\mathbb{E}[|\hat p_K(N)-p_K|]$ for $K=2,3,4$, with shaded regions indicating $\pm 1/3$ standard deviation.
  The target state is a noisy 5-qubit GHZ state with $30\%$ noise strength, 
  and the Haar-random projections are approximated by depth-5 brickwork random circuits. 
  The fully coherent protocol, $q=n=5$, requires the fewest state copies to achieve a given target error. 
  For the intermediate protocols, $q=1,\cdots,4$, the absolute error decreases with increasing $q$ at fixed copy number $N$.
  Equivalently, the number of copies $N$ required to reach a certain target error decreases as $q$ increases. 
  The fully local protocol, $q=0$, outperforms the intermediate protocols only in the small-copy regime. 
  The crossing points between the fully local and intermediate protocols shift to larger $N$ as $K$ increases. 
  This behavior reflects the fact that in the fully local protocol, measurement outcomes from each copy can be reused efficiently through collision events, 
  whereas the intermediate protocols discard rounds and restart when the measurement outcomes per copy do not coincide.}
  \label{fig:N_vs_error_all}
\end{figure*}

\begin{figure}
    \centering
    \includegraphics[width=0.7\linewidth]{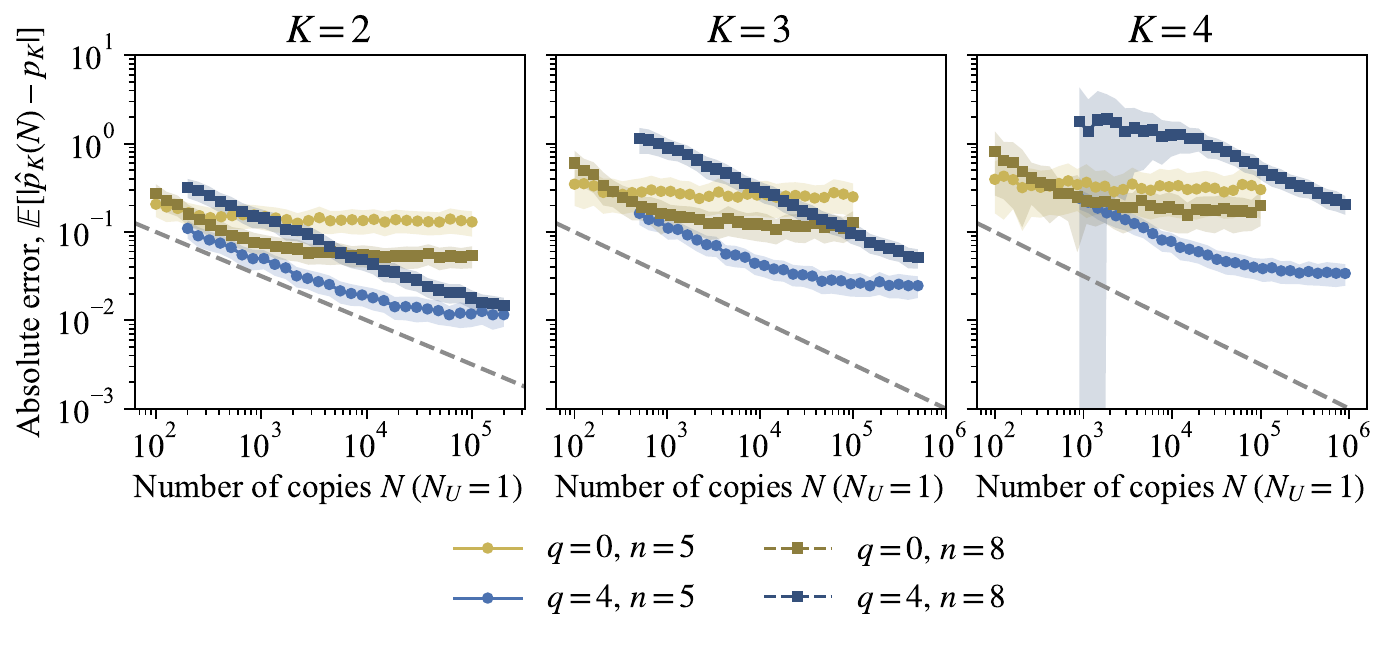}
    \caption{ 
Mean absolute estimation error 
\(\mathbb{E}[|\hat p_K(N)-p_K|]\) for \(K=2,3,4\) as a function of the total number of consumed copies \(N\), with \(N_U=1\).
The target state is a noisy GHZ state with system sizes \(n=5\) and \(n=8\), and the random unitary \(U\) is sampled from the exact Haar ensemble. 
Each curve is averaged over 500 independent repetitions, and the shaded regions indicate \(\pm 1/3\) standard deviation. 
For \(q=0\), the error initially decreases with \(N\) but eventually reaches a plateau, whose value is lower for the larger system size \(n=8\). 
For \(q=4\), increasing the total system size from \(n=5\) to \(n=8\) leads to a larger error, since the coherently measured fraction of the system becomes smaller.
}
    \label{fig:N_vs_error_diff_n_NU1}
\end{figure}

The fully local randomized measurement protocol used for comparison in this manuscript was introduced in~\cite{li2026quantum}.
This protocol estimates state moments by counting collision events among measurement outcomes.
For completeness, we restate the algorithm shown in Fig.~\ref{fig:algorithm-fully-local-protocol}.
\begin{figure}[t]
\centering
\fbox{
\begin{minipage}{0.95\columnwidth}

\textbf{Algorithm 3: Fully local protocol for estimating \(\{p_k\}_{k=2}^{K}\)~\cite{li2026quantum}}

\vspace{0.5em}

\begin{algorithmic}[1]
\Require Integers $K\geqslant2$, $N_UN_M$ copies of $\rho$ with dimension $d$
\Ensure Estimates \(\hat p_2,\cdots,\hat p_K\)

\For{\(s=1,\cdots,N_U\)}
    \State Randomly choose a unitary \(U_s\) from the (approximated) Haar-random unitary ensemble \(\mathcal E\).

    \For{\(j=1,\cdots,N_M\)}
        \State Apply \(U_s\) to \(\rho\) and measure in the computational basis.
        \State Record the outcome \(b_j\in\{0,1,\cdots,d-1\}\).
    \EndFor

    \State Let \(\mathbf b_{U_s}=(b_1,\cdots,b_{N_M})\).

    \For{\(k=2,\cdots,K\)}
        \State Compute
        \[
        \hat M_k^{U_s}
        =
        \frac{\binom{k+d-1}{k}}
        {d\binom{N_M}{k}}
        \sum_{i_1<\cdots<i_k}
        \Ind{
        b_{i_1}=\cdots=b_{i_k}
        } .
        \]
    \EndFor
\EndFor

\For{\(k=2,\cdots,K\)}
    \State Compute
    \[
    \hat \zeta_k
    =
    \frac{1}{N_U}
    \sum_{s=1}^{N_U}
    \hat M_k^{U_s}.
    \]
\EndFor

\State Substitute \(\hat\zeta_2,\cdots,\hat\zeta_K\) into the moment-polynomial relations.

\State Solve sequentially for \(\hat p_2,\cdots,\hat p_K\).

\State \Return \(\hat p_2,\cdots,\hat p_K\).

\end{algorithmic}

\end{minipage}
}
\caption{
Fully local randomized measurement protocol for estimating $p_2,\cdots,p_K$.
}
\label{fig:algorithm-fully-local-protocol}
\end{figure}
In this algorithm, the moment-polynomial relation between $\zeta_k$ and $p_k$ is:
\begin{equation}
    \zeta_k = \frac{1}{k!}\sum_{\pi \in S_k}\tr(V_{\pi} \rho^{\otimes k}) = \frac{1}{k!}\sum_{\pi \in S_k}\prod_{j=1}^{c(\pi)}p_{\nu_j}.
\end{equation}
The comparisons between the fully local protocol and our protocol are shown in Fig.~\ref{fig:N_vs_error_all} and Fig.~\ref{fig:N_vs_error_diff_n_NU1}.
In addition to the behavior discussed in the main text, Fig.~\ref{fig:N_vs_error_all} shows that the range of $N$ over which the fully local protocol outperforms the intermediate projected protocols becomes wider as $K$ increases.
This is because the fully local protocol uses all measurement outcomes through collision counting, 
whereas our protocol incurs a postselection overhead whenever the measured projection outcomes do not coincide.
This overhead becomes more severe for larger $K$, since the coincidence probability decreases with the number of copies.
In Fig. \ref{fig:N_vs_error_diff_n_NU1}, we fix $N_U=1$ and compare different system sizes.
For the fully local protocol, the error becomes smaller when $n$ increases. 
By contrast, for our protocol with fixed $q=4$, the error becomes larger as the full system size increases, 
because the compression becomes more aggressive as the retained fraction $q/n$ decreases.
Nevertheless, the projected protocol can still outperform the fully local protocol once $N$ is sufficiently large.

\end{document}